\documentclass[twocolumn,showpacs,preprintnumbers,amsmath,amssymb,prl,superscriptaddress,longbibliography]{revtex4-2}
\usepackage{bbm}
\usepackage{mathrsfs}
\usepackage{graphicx}
\usepackage{dcolumn}
\usepackage{bm}
\usepackage{amsmath}
\usepackage{amsfonts}
\usepackage[dvipsnames]{xcolor}
\usepackage[colorlinks=true,linkcolor=Blue,urlcolor=BlueViolet,citecolor=BlueViolet]{hyperref}
\usepackage{floatrow}

\begin{document}

\title{Moir\'e Ferroelectricity–Driven Band Engineering in Twisted Square Bilayers}
\author{Kejie Bao}
\affiliation{State Key Laboratory of Surface Physics and Department of Physics, Fudan University, Shanghai 200433, China}
\affiliation{Shanghai Research Center for Quantum Sciences, Shanghai 201315, China}
\author{Rui Shi}
\affiliation{State Key Laboratory of Surface Physics and Department of Physics, Fudan University, Shanghai 200433, China}
\affiliation{Shanghai Research Center for Quantum Sciences, Shanghai 201315, China}
\author{Huan Wang}
\affiliation{State Key Laboratory of Surface Physics and Department of Physics, Fudan University, Shanghai 200433, China}
\affiliation{Shanghai Research Center for Quantum Sciences, Shanghai 201315, China}
\author{Linghao Huang}
\affiliation{State Key Laboratory of Surface Physics and Department of Physics, Fudan University, Shanghai 200433, China}
\affiliation{Shanghai Research Center for Quantum Sciences, Shanghai 201315, China}
\author{Jing Wang}
\thanks{wjingphys@fudan.edu.cn}
\affiliation{State Key Laboratory of Surface Physics and Department of Physics, Fudan University, Shanghai 200433, China}
\affiliation{Shanghai Research Center for Quantum Sciences, Shanghai 201315, China}
\affiliation{Institute for Nanoelectronic Devices and Quantum Computing, Fudan University, Shanghai 200433, China}
\affiliation{Hefei National Laboratory, Hefei 230088, China}

\begin{abstract}

We develop the moir\'e band theory for M-valley twisted square homobilayers with layer groups $P$-$42m$ and $P$-$4m2$, and propose candidate material realizations. We show that moir\'e ferroelectricity—originating from sliding ferroelectricity in the untwisted bilayers—provides an independent control knob for miniband engineering in addition to interlayer tunneling. The competition between these two effects enables controlled switching between layer-resolved bilayer minibands and an effective single isolated miniband. Remarkably, these systems exhibit an emergent momentum-space nonsymmorphic symmetry in the absence of external magnetic fields. Large-scale \emph{ab initio} calculations identify Cu$_2$WS$_4$ and GeCl$_2$ as representative materials realizing the ferroelectricity- and tunneling-dominated regimes, respectively. Our results establish twisted square homobilayers as a promising platform for correlated band engineering beyond moir\'e hexagonal systems.
\end{abstract}


\maketitle

The emergence of twisted van der Waals materials has established moir\'e superlattices as a versatile for exploring strongly correlated and topological quantum phases~\cite{Andrei2021,Kennes2021,balents2020,carr2020,Mak2022}. Landmark discoveries—including fractional Chern insulators~\cite{Cai2023,Zeng2023,Park2023,Xu2023,Lu2024,li2021,devakul2021,Wang2024,Jia2024,Yu2024,macdonald2024} and unconventional superconductivity~\cite{Cao2018_2,yankowitz2019,Xia2024,Guo2025,knuppel2025correlated}—have predominantly emerged in hexagonal lattice such as graphene and transition metal dichalcogenides (TMD)~\cite{Wu2018,Wu2019_tmd,Bistritzer2011,Angeli2021,Calugaru2025,Bao2025,Lei2025}. Recently, increasing attention has turned to realizing square lattice moir\'e systems~\cite{kariyado2019,luo2021,Eugenio2025,Kariyado2025b,xu2025engineering,soeda2022,Li2022magic,sarkar2025,can2021high,zhao2023time,song2022doping,volkov2023current,eugenio2023twisted,kariyado2026moir,twistgamma2026}. Such systems could provide a highly tunable platform for simulating the Hubbard model relevant to cuprate~\cite{scalapino2012,arovas2022hubbard,qin2022hubbard} and nickelate superconductors~\cite{Qu2024bilayer,Fan2024super,kaneko2024pair,lu2024interlayer,jiang2025theory}. 
Despite several theoretical proposals for twisted square lattice systems, most existing approaches generate moir\'e minibands primarily through interlayer tunneling~\cite{Eugenio2025,Kariyado2025b,xu2025engineering}. Consequently, the tunability of the low-energy electronic structure remains limited. A key open question is therefore whether symmetry-based mechanisms can provide additional control knobs to tune both the number and layer polarization of low-energy minibands. Achieving such control would significantly expand the design principles of twistronics and enable the realization of a broader class of correlated Hamiltonians.

Here we identify such a control knob in twisted square lattice homobilayers: moir\'e ferroelectricity (FE). It produces a symmetry-selected, layer-asymmetric moir\'e potential, directly competing with interlayer tunneling  and qualitatively reshaping the low-energy minibands. FE  has emerged as an efficient band engineering mechanism in twisted TMD~\cite{zhang2024polarization}. Microscopically, moir\'e FE originates from sliding FE in aligned bilayers, where stacking-dependent interlayer charge transfer generates an out-of-plane polarization (OP) and thus a layer-asymmetric electrostatic potential between the two layers~\cite{li2017binary,yasuda2021stacking,vizner2021interfacial,fei2018ferroelectric,wang2022interfacial,Wan2022room,ji2023general,liu2023atypical}. In the presence of a small twist angle, this potential acquires moir\'e periodicity, giving rise to moir\'e FE. The resulting competition between the FE layer-asymmetric potential and interlayer tunneling enables controlled switching between a single-band Hubbard model and a layer-resolved bilayer Hubbard model.

In this Letter, we develop a symmetry-based framework for moir\'e FE in twisted square lattice homobilayers, focusing on the layer group $P$-$42m$ and $P$-$4m2$. For systems with band extrema at M point of the Brillouin zone (BZ), we show that moir\'e FE controls the number, isolation, and layer polarization of the low-energy minibands. The effective tight-binding descriptions are constructed in both the FE- and tunneling-dominated limits. We further uncover an intrinsic momentum space nonsymmorphic symmetry that emerges in these systems without external magnetic flux. Finally, large-scale density functional theory (DFT) calculations identify twisted Cu$_2$WS$_4$ and GeCl$_2$ as representative  materials realizing the two limits.

\begin{figure}[t]
  \begin{center}
    \includegraphics[width=3.4in,clip=true]{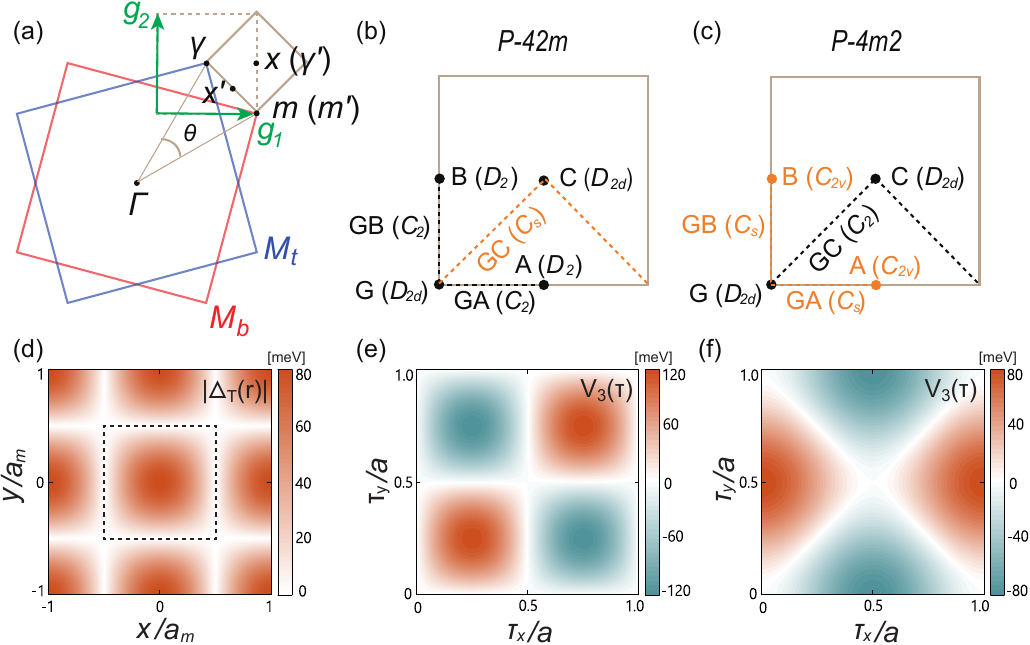}
      \caption{(a) Brillouin zones (BZ) of the top and bottom layers, shown as red and blue squares, respectively. The dashed (solid) gray square denotes the moir\'e BZ used in the continuum model (DFT calculations), where the high-symmetry points are marked by Greek letter without (with) prime. (b,c) OP and corresponding bilayer point groups for $\{E|\tau\}$ stacking in the layer group $P$-$42m$ and $P$-$4m2$, respectively. Orange lines indicate the presence of OP. Translation vectors G(0,0), C(0.5,0.5), A(0.5,0), and B(0,0.5) are labeled. (d) Spatial distribution of $|\Delta_{T}(\mathbf{r})|$ with $w_0=20$~meV. 
      Spatial patterns of the lowest-harmonic FE potential for (e) $P$-$42m$ with $w^{(2)}_1=-w^{(2)}_2=30$~meV, and (f) $P$-$4m2$ with $w^{(1)}_1=-w^{(1)}_2=-20$~meV.}
      \label{fig1}
  \end{center}
\end{figure} 

\emph{Moir\'e band theory---}We begin with the sliding FE of untwisted square lattice homobilayers with a relative shift $\boldsymbol{\tau}$, focusing on two tetragonal layer group $P$-$42m$ and $P$-$4m2$. Symmetry analysis shows that OP is allowed in both of them with the stacking configuration $\{E|\tau\}$, that transforms the bottom layer to the top layer~\cite{ji2023general,supple}. Here $E$ denotes the identity, and $\tau$ the translation vector. Figs.~\ref{fig1}(b) and~\ref{fig1}(c) show the characteristic spatial distributions of OP in $P$-$42m$ and $P$-$4m2$, respectively. In $P$-$42m$, OP emerges along the line GC, which is enforced by the mirror symmetries $m_{110}$ or $m_{1\bar{1}0}$. $m_{110}$ denotes the mirror reflection through the $(110)$ plane. In contrast, for $P$-$4m2$, OP remains symmetry-allowed along the lines GA and GB by $m_{100}$ and $m_{010}$, respectively. Specifically, OP at points A and B is enforced by $2_{001}$, the twofold rotation about the $[001]$ axis. As shown below, the resulting spatially varying OP acts as a periodic potential that strongly modulates the moir\'e minibands near the BZ corner.

To elucidate band engineering with moir\'e FE, we construct the continuum model. Without loss of generality, suppose that the valence band maximum (VBM) locates at the BZ corner (M point) and the spin-orbit coupling is negligible, then the spinless two-band $\mathbf{k}\cdot\mathbf{p}$ Hamiltonian for the untwisted bilayer is
\begin{equation}\label{eq1}
	\mathcal{H}_{\text{un}} = \begin{pmatrix}
		\frac{\mathbf{k}^2}{2m} + V_{t}(\boldsymbol{\tau}) & \Delta_T(\boldsymbol{\tau})\\
		\Delta_T^\dagger(\boldsymbol{\tau}) & \frac{\mathbf{k}^2}{2m} + V_{b}(\boldsymbol{\tau})
	\end{pmatrix}\equiv\frac{\mathbf{k}^2}{2m}+\sum\limits_{i=0}^3V_i(\boldsymbol{\tau})\sigma_i,\nonumber
\end{equation}
where $m$ is the effective mass around M point. $V_{t/b}$ the layer-dependent potential and $\Delta_T$ the interlayer tunneling, are periodic functions of $\boldsymbol{\tau}$ and can be Fourier expanded, with coefficients constrained by the time-reversal symmetry $\mathcal{T}$ and crystalline symmetry of the bilayer~\cite{supple,Akashi2017}. $(\sigma_0,\sigma_1,\sigma_2,\sigma_3)$ are Pauli matrices acting on layer. The layer-asymmetric term $V_3$, encoding the interlayer potential difference generated by FE polarization, can strongly reshape the moir\'e band structures in twisted homobilayer. For $P$-$42m$, the symmetry-allowed lowest-harmonic form is 
$V_{3}(\boldsymbol{\tau})=(w^{(2)}_{1} - w^{(2)}_{2})(\text{cos}(\tilde{\tau}_x-\tilde{\tau}_y) - \text{cos}(\tilde{\tau}_x+\tilde{\tau}_y))$, where $\tilde{\tau}_\mu = 2\pi \tau_\mu/a$ and $a$ is the lattice constant. The typical spatial profile of $V_{3}(\boldsymbol{\tau})$ is shown in Fig.~\ref{fig1}(e), consistent with the symmetry analysis in Fig.~\ref{fig1}(b). While for $P$-$4m2$, the lowest-harmonic of $V_3(\boldsymbol{\tau})=(w^{(1)}_1-w^{(1)}_2)(\text{cos}\ \tilde{\tau}_y-\text{cos}\ \tilde{\tau}_x)$ and its spatial pattern is displayed in Fig.~\ref{fig1}(f).

For the twisted homobilayer with angle $\theta$, the moir\'e potentials are obtained by substituting the local shift $\boldsymbol{\tau}=2\sin(\theta/2)\hat{\mathbf{z}}\times\mathbf{r}$, which periodically varies with the moir\'e lattice vectors $\mathbf{L}_i=\mathbf{a}_i\times\hat{\mathbf{z}}/[2\sin(\theta/2)]$. Then the effective continuum Hamiltonian for the twisted homobilayer is~\cite{supple},
\begin{equation}\label{eq2}
	\mathcal{H}_{\text{twisted}} = \begin{pmatrix}
		\frac{(\mathbf{k} - \mathbf{M}_t)^2}{2m} + V_{t}(\mathbf{r}) & \Delta_T(\mathbf{r})\\
		\Delta_T^\dagger(\mathbf{r}) & \frac{(\mathbf{k} - \mathbf{M}_b)^2}{2m} + V_{b}(\mathbf{r})
	\end{pmatrix}.
\end{equation}
The kinetic energies in the diagonal terms are centered at the M-point momenta $\mathbf{M}_{t/b}$ in the top and bottom layers, respectively. In the small $\theta$ limit, they scale as $1/|\mathbf{L}_i^2|$ and become less important, and the low-energy minibands of Eq.~(\ref{eq2}) are governed by the competition between the layer-dependent  potential and interlayer tunneling. As we show below, in the moir\'e FE-dominated regime, the minibands become layer resolved and can be naturally described by coupled bilayer lattice models. In the opposite tunneling-dominated regime, interlayer hybridization reconstructs the spectrum into a single isolated band.

\emph{Twisted $P$-$42m$---}We first consider the layer group $P$-$42m$. By invoking the $C_{2z}$, $C_{2y}$ and $\mathcal{T}$ symmetries of the M valley in the twisted bilayer, the moir\'e potentials take the following lowest-harmonic form:
\begin{eqnarray}\label{eq3}
        V_{t}(\mathbf{r}) &=& 2w^{(1)}_{1}\cos(\mathbf{g}_1\cdot\mathbf{r}) + 2w^{(2)}_{1}\cos\left((\mathbf{g}_1 + \mathbf{g}_2)\cdot\mathbf{r}\right)
        \nonumber
        \\
        &&+ 2w^{(1)}_{2}\text{cos}(\mathbf{g}_2\cdot\mathbf{r}) + 2w^{(2)}_{2}\cos((\mathbf{g}_1 - \mathbf{g}_2)\cdot\mathbf{r}),
        \nonumber
        \\
        \Delta_{T}(\mathbf{r}) &=& w_0 \big( 1 + e^{i\mathbf{g}_1\cdot\mathbf{r}} + e^{i\mathbf{g}_2\cdot\mathbf{r}} + e^{i(\mathbf{g}_1 + \mathbf{g}_2)\cdot\mathbf{r}}\big),\nonumber
\end{eqnarray}
where $\mathbf{g}_{i}$ are the moir\'e reciprocal lattice vectors satisfying $\mathbf{g}_{i}\cdot\mathbf{L}_{j}= 2\pi\delta_{ij}$ and $\mathbf{k}_0=\mathbf{M}_t - \mathbf{M}_b=(\mathbf{g}_1 + \mathbf{g}_2)/2$. $V_b(\mathbf{r})=V_{t}(C_{2y}\mathbf{r})$. 

\begin{figure}[t]
  \begin{center}
    \includegraphics[width=3.4in,clip=true]{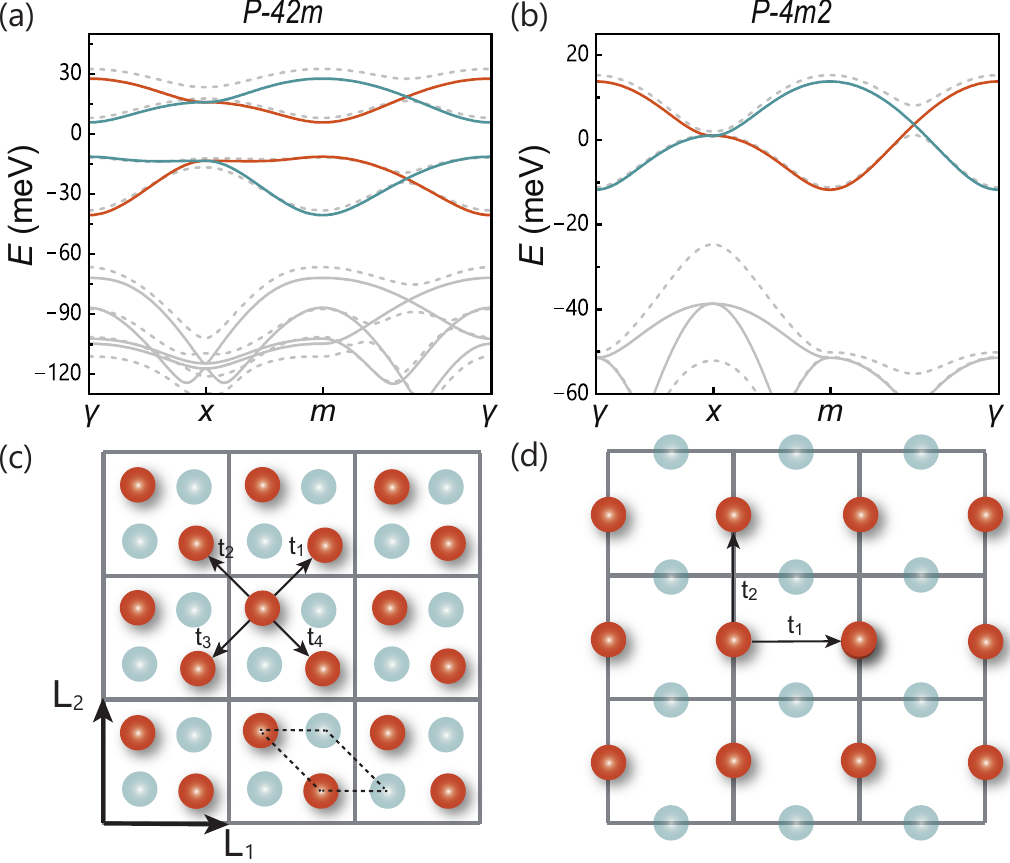}
      \caption{Band structure calculated from the continuum model for (a) $P$-$42m$ and (b) $P$-$4m2$, where FE potential dominates over interlayer tunneling. Solid (dashed) lines denote bands without (with) interlayer tunneling. The parameters are $m=0.6m_0$ ($m_0$ the free electron mass), $\theta=6^\circ$, $a=5$~\AA. For (a): $w^{(1)}_1=-w^{(1)}_2=20$~meV, $w^{(2)}_1=-w^{(2)}_2=30$~meV, $w_0=10$~meV. For (b): $w^{(1)}_1=-w^{(1)}_2=-20$~meV, $w_0=6$~meV, $w'_0=5$~meV. The low-energy bands of the top (bottom) layer in the absence of interlayer tunneling are highlighted in brown (jasper). The corresponding tight-binding models are schematically illustrated in (c) and (d) using the same color scheme, where $t_i$ denote the dominant hopping amplitudes for the top layer.}
      \label{fig2}
  \end{center}
\end{figure}

The moir\'e band structure is shown in Fig.~\ref{fig2}(a). To see how moir\'e FE affects the band structures, we first study the FE-dominated limit, where interlayer tunneling can be treated perturbatively. The two layers then decouple, and the low-energy physics is governed by the intralayer potential $V_{t/b}(\mathbf{r})$. For the top layer, $V_t(\mathbf{r})$ exhibits two maxima per moir\'e unit cell near $(-1/4, 1/4)$ and $(1/4, -1/4)$, which trap the holes~\cite{supple}. The resulting low-energy states are therefore described by a two-orbital tight-binding model, consistent with the continuum band structure in Fig.~\ref{fig2}(a). Extracting only nearest neighbor hopping parameters from the continuum model yields dominant anisotropic hoppings $t_1>t_{2,4}>t_{3}$ [Fig.~\ref{fig2}(c)], which corresponds to an extended two-dimensional Su–Schrieffer–Heeger (SSH) model consisting of coupled SSH chains along the diagonal direction. The tight-binding model of the bottom layer is related to that of the top layer by $C_{2y}$ operation. Then including interlayer tunneling opens a hybridization gap, which is shown as the dashed line in Fig.~\ref{fig2}(a), and the system maps onto a bilayer extended SSH model with mutually perpendicular SSH chains in the two layers. In the opposite limit where interlayer tunneling dominates, low-energy physics is thus governed by the spatial profile of $|\Delta_T(\mathbf{r})|$  [Fig.~\ref{fig1}(d)], leading to an effective single-band model analogous to that in Ref.~\cite{Eugenio2025}. The explicit forms of the tight-binding model are in Supplemental Material~\cite{supple}.

\emph{Twisted $P$-$4m2$---}We next study layer group 
$P$-$4m2$, where the twisted bilayer of the M valley preserves $\mathcal{T}$, $C_{2z}$ and a diagonal twofold rotation $C_{2d}$ symmetries. The moir\'e potentials take the following lowest-harmonic form:
\begin{eqnarray}\label{eq4}
        V_{t}(\mathbf{r}) &=& 2w^{(1)}_{1}\text{cos}(\mathbf{g}_1\cdot\mathbf{r}) + 2w^{(1)}_{2}\text{cos}(\mathbf{g}_2\cdot\mathbf{r}),
        \nonumber
        \\
        \Delta_{T}(\mathbf{r}) &=& w_0\big(1 + e^{i(\mathbf{g}_1 + \mathbf{g}_2)\cdot\mathbf{r}}\big) + w'_0 \big( e^{i\mathbf{g}_1\cdot\mathbf{r}} + e^{i\mathbf{g}_2\cdot\mathbf{r}}\big),\nonumber
\end{eqnarray}
and $V_b(\mathbf{r})=V_{t}(C_{2d}\mathbf{r})$. The moir\'e band structure is calculated in Fig.~\ref{fig2}(b), where FE dominates over the interlayer tunneling. To analyze this, we first neglect the interlayer tunneling and study the spatial profile of $V_{t/b}(\mathbf{r})$. For the top layer, $V_{t}(\mathbf{r})$ exhibits a single maximum per moir\'e unit cell at $(1/2, 0)$. The low-energy physics of the top layer is therefore captured by a single orbital on a square lattice, with anisotropic nearest-neighbor hoppings in the corresponding tight-binding model, consistent with the miniband structure shown in Fig.~\ref{fig2}(b).  While the model of bottom layer is related to that of top layer by $C_{2d}$. The interlayer tunneling opens a hybridization gap between the top- and bottom-layer bands, shown as the dashed line in Fig.~\ref{fig2}(d). Therefore, the whole system is described by two nested square lattices displaced along the out-of-plane direction [Fig.~\ref{fig2}(d)]. In the limit where $\Delta_{T}$ dominates over FE potential, the system again reduces to an effective single-band model on square lattice~\cite{supple}.

\emph{Emergent nonsymmorphic symmetry---}It is intriguing to point out that a momentum space nonsymmorphic symmetry emerges in these systems. The single particle basis for the Hamiltonian in Eq.~\eqref{eq2} is $\hat{c}^{\dagger}_{\mathbf{k},\mathbf{Q},l} = \hat{c}^{\dagger}_{\mathbf{M}_l+\mathbf{k}-\mathbf{Q}_l,l}$,  where $\mathbf{Q}_l\in$\{$\delta\mathbf{q}_l + m\mathbf{g}_1+n\mathbf{g}_2|m,n\in \mathbb
Z$\}, with $\delta\mathbf{q}_l = 0$ for $l=t$ and $\delta\mathbf{q}_l = \mathbf{k}_0$ for $l=b$. For the layer group $P$-$42m$, the action of $C_{2y}$ on the creation operator reads
\begin{equation}
C_{2y}\hat{c}^{\dagger}_{\mathbf{k},\mathbf{Q},l}C_{2y}^{-1} = \hat{c}^{\dagger}_{C_{2y}\mathbf{k}+\mathbf{k}_0,C_{2y}\mathbf{Q}+\mathbf{k}_0,\bar{l}},
\end{equation}
where $\bar{l}=t\ (b)$ for $l=b\ (t)$. Recalling $\mathbf{k}_0=(\mathbf{g}_1+\mathbf{g}_2)/2$, thus, the in-plane rotation $C_{2y}$ acts as a nonsymmorphic operation in momentum space, an intrinsic property that does not rely on external magnetic flux. This behavior originates from a layer-dependent vector potential $\mathbf{A}_l = (l/4e)(\mathbf{g}_1-\mathbf{g}_2)$ arising from the relative M-valley shift between the two layers~\cite{supple,Calugaru2025}. In this gauge field picture, an electron traversing the dashed loop in Fig.~\ref{fig2}(c), which connects the two layers, acquires a $\pi$ flux. Although this flux preserves $\mathcal{T}$ symmetry, it nontrivially modifies certain hopping amplitudes of the bottom layer tight-binding model by an additional minus sign when applying $C_{2y}$ operation to top layer model~\cite{supple}, which reflects the projective representation formed by $C_{2y}$ and translation symmetries~\cite{Chen2022,Zhang2023}. In contrast, for the layer group $P$-$4m2$, the momentum space nonsymmorphic action associated with $C_{2d}$ can be removed by a gauge transformation that shifts center of the moir\'e BZ from $\gamma$ to $x$ point~\cite{supple}. Consequently, $C_{2d}$ becomes a symmorphic symmetry operation, and the bottom-layer tight-binding model follows directly from applying 
$C_{2d}$ to the top-layer model.

\begin{figure}[t]
  \begin{center}
    \includegraphics[width=3.4in,clip=true]{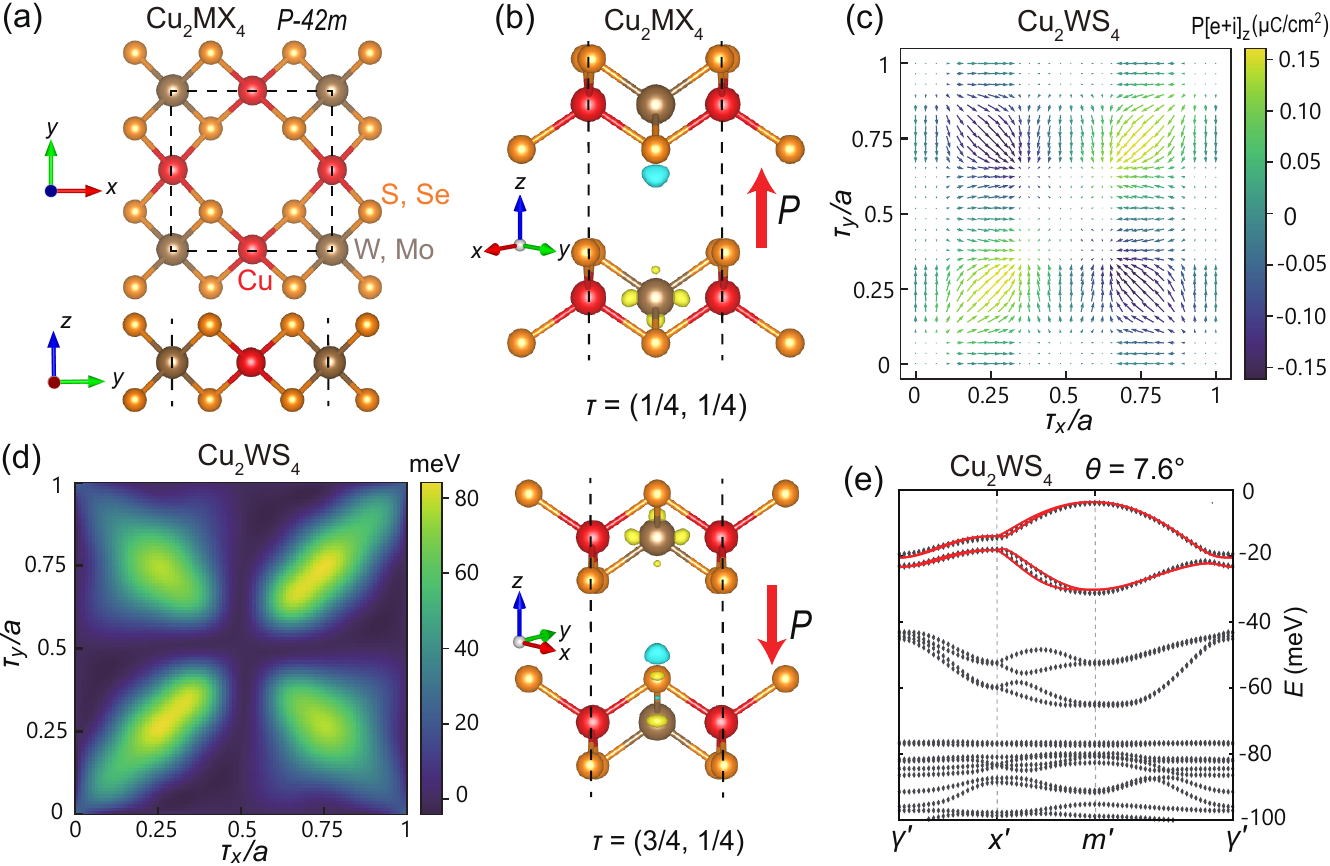}
      \caption{(a) Atomic structure of monolayer Cu$_2$\emph{MX}$_4$ from top and side views. (b) Differential charge-density distributions and OP directions for $\{E|\tau\}$-stacked bilayers with $\bm{\tau}=(1/4,1/4)$ and $(3/4,1/4)$. Yellow and blue isosurfaces denote electron accumulation and depletion upon stacking, respectively. (c) OP distribution for different stacking configurations; the arrow direction indicates the in-plane polarization component, while the color encodes the out-of-plane component. (d) Band splitting at the M point as a function of stacking. (e) Band structures of twisted bilayer Cu$_2$WS$_4$ at $\theta=7.6^\circ$. Gray dots show DFT results, and red lines denote the fitting from the continuum Hamiltonian.}
      \label{fig3}
  \end{center}
\end{figure}

\emph{Candidate materials---}The monolayer Cu$_2$\emph{MX}$_4$ ($M$ = W, Mo; $X$ = S, Se) crystallizes in a square lattice with layer group $P$-$42m$ and have been experimentally synthesized~\cite{pruss1993new,crossland2005synthesis,chen2014solvothermal,lin2019recent}. As shown in Fig.~\ref{fig3}(a), each Cu and W/Mo atom is coordinated by four S/Se atoms in a distorted edge-sharing tetrahedron. VBM of the monolayer lies at the M point with the effective mass $0.6m_0$, and is predominantly contributed by the antibonding Cu $d_{xz,yz}$ orbitals~\cite{supple}, For bilayer Cu$_2$WS$_4$ with $\{E|\tau\}$ stacking, the calculated polarization distribution agrees with the symmetry analysis and exhibits four extrema at $\bm{\tau}=(1/2\pm1/4,1/2\pm1/4)$. As shown in Fig.~\ref{fig3}(c), the polarization points toward the top layer for $\bm{\tau}=(1/4,1/4)$ and $(3/4,3/4)$, whereas for $\bm{\tau}=(3/4,1/4)$ and $(1/4,3/4)$ it points toward the bottom layer. DFT calculations yield an OP of approximately $0.16$~$\mu\mathrm{C}/\mathrm{cm}^2$, originating from charge transfer between S atoms in the top (bottom) layer and W atoms in the bottom (top) layer [Fig.~\ref{fig3}(b)]. To further quantify the FE, we compute the band splitting of the untwisted bilayer at the M point, $\Delta E_M=2\sqrt{V^2_1(\boldsymbol{\tau})+V^2_3(\boldsymbol{\tau})}$, as a function of the relative shift $\boldsymbol{\tau}$. The resulting splitting closely follows the polarization pattern in Fig.~\ref{fig3}(d), indicating that the layer-asymmetric potential dominates over interlayer tunneling in Cu$_2$WS$_4$.
 
We next perform large-scale \emph{ab initio} calculations for twisted bilayer. Different from the continuum model, the minimal commensurate supercell used in DFT is defined by $\tilde{\mathbf{L}}_{1}=(n+1)\mathbf{a}_{1}+n\mathbf{a}_{2}$ and $\tilde{\mathbf{L}}_{2}=-n\mathbf{a}_{1}+(n+1)\mathbf{a}_{2}$, where $n$ is an integer and $2\sin(\theta/2)=1/\sqrt{n^{2}+n+1/2}$. Each supercell therefore contains two effective moir\'e unit cells of the continuum model~\cite{kariyado2019,supple}. For $\theta=7.6^\circ$, the calculated band structure shown in Fig.~\ref{fig3}(e) exhibits two isolated groups of bands, each consisting of a pair of doubly degenerate bands that are well separated from the remaining states. This double degeneracy arises from effective band folding due to the enlarged supercell in DFT calculations. As the twist angle decreases, the bands become increasingly flat, leaving only the uppermost group of bands near the Fermi energy.

Our continuum model with parameters $w^{(1)}_1=17$~meV, $w^{(1)}_2=-11$~meV, $w^{(2)}_1=44$~meV, $w^{(2)}_2=-66$~meV and $w_0=3$~meV accurately reproduces the topmost eight valence bands of twisted Cu$_2$WS$_4$. In particular, the exact degeneracy along the high symmetry path $m'$-$\gamma'$ is protected by $C_{2y}$. The hierarchy $|w^{(2)}_i|\gg w_0$ confirms that the FE potential dominates over interlayer tunneling, consistent with the analysis of the M point splitting. Moreover, the extracted parameters are comparable to those used in the model analysis, indicating that the two isolated groups of low-energy bands are well described by the tight-binding model shown in Fig.~\ref{fig2}(c). 

\begin{figure}[t]
  \begin{center}
    \includegraphics[width=3.4in,clip=true]{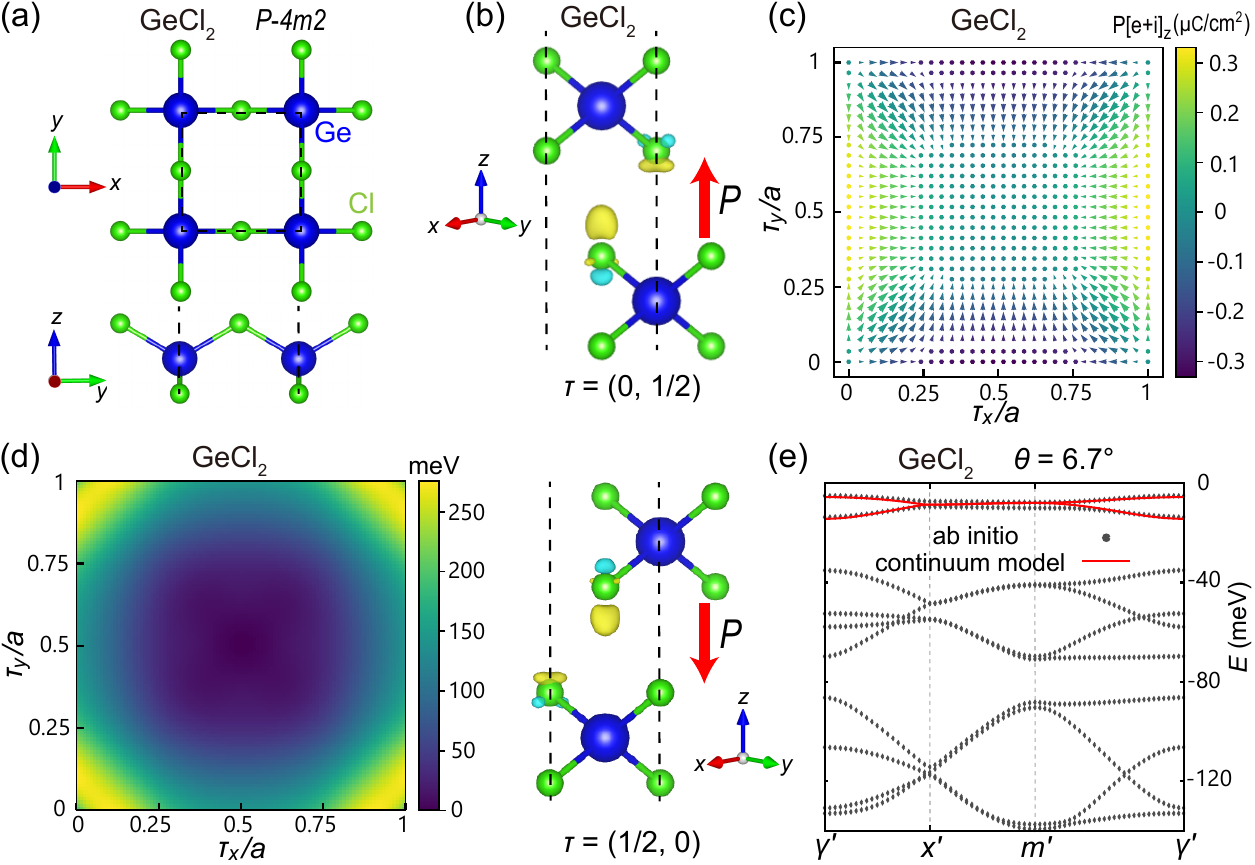}
      \caption{(a) Atomic structure of monolayer GeCl$_2$.(b) Differential charge-density distributions and OP directions for $\{E|\tau\}$-stacked bilayers with $\bm{\tau}=(0,1/2)$ and $(1/2,0)$. (c) OP distributions for different stacking configurations. (d) Band splitting at the M point as a function of stacking. (e) Band structures of twisted bilayer GeCl$_2$ at $\theta=6.7^\circ$ by DFT calculations and continuum model. The style follows that of  Fig.~\ref{fig3}.}
      \label{fig4}
  \end{center}
\end{figure} 

GeCl$_2$ is another candidate material with layer group $P$-$4m2$. As shown in Fig.~\ref{fig4}(a), each Ge atom is tetrahedrally coordinated by four Cl atoms, and neighboring tetrahedra are connected via corner sharing~\cite{jiang20242d}. VBM lies at the M point with effective mass $1.2m_0$ and is predominately contributed by the anti-bonding Cl $p_x,p_y$ orbitals~\cite{supple}. For the $\{E|\tau\}$ stacking, the polarization extrema appear at the A and B points with opposite orientations, consistent with the symmetry analysis [Fig.~\ref{fig4}(c)]. The maximum OP reaches approximately $0.33$~$\mu\mathrm{C}/\mathrm{cm}^2$, originating from charge accumulation on Cl atoms in the bottom (top) layer [Fig.~\ref{fig4}(b)]. The band splitting at M point in Fig.~\ref{fig4}(d) reflects the combined effects of interlayer tunneling and FE. Fig.~\ref{fig4}(e) shows the DFT calculations for twisted GeCl$_2$ at $\theta=6.7^\circ$. In contrast to Cu$_2$WS$_4$, only a single nearly degenerate band—arising from band folding—is well isolated from the remaining bands. This behavior reflects the comparable magnitudes of the FE potential and interlayer tunneling, as indicated by the fitted parameters $w^{(1)}_1=-w^{(1)}_2=-20$~meV, $w_0=w'_0=30$~meV. Nevertheless, owing to the larger effective mass and slightly stronger interlayer tunneling, twisted GeCl$_2$ still hosts an isolated flat band. The exact degeneracy along the $x'$-$m'$ is protected by the $C_{2d}$ symmetry. Moreover, the system retains an effective $C_{4z}$ symmetry for this parameter set~\cite{supple}. 

\emph{Discussions---}We have developed a moir\'e band theory for twisted square homobilayers and identified candidate material platforms in which FE and interlayer tunneling compete to determine low-energy bands. The FE patterns arising from the $\{E|\tau\}$ stacking in the $P$-$42m$ and $P$-$4m2$ layer groups are analyzed. More generally, the FE pattern can be engineered through the monolayer symmetry and the generalized stacking operator $\hat{O}$, while its magnitude can be tuned by modifying the interlayer charge transfer via elemental substitution~\cite{supple}. Moir\'e FE therefore provides an additional control knob for miniband engineering beyond interlayer tunneling.

The layer-resolved minibands in twisted Cu$_2$WS$_4$, arising from the dominant moir\'e FE potential, may serve as a tunable platform for realizing the bilayer Hubbard model, which hosts a variety of correlated phenomena including magnetism, metal–insulator transitions, and unconventional superconductivity~\cite{scalapino2012,arovas2022hubbard}. 
Near half filling of the topmost valence miniband, hole pockets appear at $\gamma$ and $m$, originating primarily from orbitals on the top and bottom layers, respectively. These pockets coincide under $(\pi,\pi)$ translation by symmetry. Such a Fermi-surface geometry can enhance inter-pocket scattering at $Q\approx(\pi,\pi)$, potentially promoting spin-density-wave instabilities and antiferromagnetic order~\cite{bouadim2008,mou2022}. Spin fluctuations may further mediate competing pairing channels, including inter-pocket sign-changing ($s^\pm$) pairing or intra-pocket $d$-wave pairing, with the dominant channel determined by the competition between intra- and inter-orbital scattering~\cite{bulut1992,maier2011}.
In addition, the in-plane hopping anisotropy may further favor stripe order~\cite{normand2001}.

Our theory for twisted M-valley of the square lattice can also be generalized to rectangular lattice with layer groups $P222$ or $C222$. In the $P$-$42m$ case, the $S_{4z}$ symmetry is explicitly broken after twisting, leaving only $C_{2z}$, which leads to the same effective space group as in the rectangular lattice. Meanwhile, the momentum space nonsymmorphic symmetry $C_{2y}$ remains preserved in twisted rectangular M-valley systems. Recent work suggests that such symmetries can give rise to quantized crystalline electromagnetic responses in insulating states~\cite{Vaidya2025}. Twisted square 
M-valley systems therefore provide an experimentally accessible platform, complementary to twisted hexagonal systems, for realizing and probing such emergent symmetry.

\begin{acknowledgments}
\emph{Acknowledgement---}This work is supported by the Natural Science Foundation of China through Grant No.~12350404, National Key Research Program of China under Grant No.~2025YFA1411400, Quantum Science and Technology-National Science and Technology Major Project through Grant No.~2021ZD0302600, the Science and Technology Commission of Shanghai Municipality under Grants No.~23JC1400600, No.~24LZ1400100 and No.~2019SHZDZX01. K.B. and R.S. contributed equally to this work.
\end{acknowledgments}


\begin{thebibliography}{75}%
\makeatletter
\providecommand \@ifxundefined [1]{%
 \@ifx{#1\undefined}
}%
\providecommand \@ifnum [1]{%
 \ifnum #1\expandafter \@firstoftwo
 \else \expandafter \@secondoftwo
 \fi
}%
\providecommand \@ifx [1]{%
 \ifx #1\expandafter \@firstoftwo
 \else \expandafter \@secondoftwo
 \fi
}%
\providecommand \natexlab [1]{#1}%
\providecommand \enquote  [1]{``#1''}%
\providecommand \bibnamefont  [1]{#1}%
\providecommand \bibfnamefont [1]{#1}%
\providecommand \citenamefont [1]{#1}%
\providecommand \href@noop [0]{\@secondoftwo}%
\providecommand \href [0]{\begingroup \@sanitize@url \@href}%
\providecommand \@href[1]{\@@startlink{#1}\@@href}%
\providecommand \@@href[1]{\endgroup#1\@@endlink}%
\providecommand \@sanitize@url [0]{\catcode `\\12\catcode `\$12\catcode `\&12\catcode `\#12\catcode `\^12\catcode `\_12\catcode `\%12\relax}%
\providecommand \@@startlink[1]{}%
\providecommand \@@endlink[0]{}%
\providecommand \url  [0]{\begingroup\@sanitize@url \@url }%
\providecommand \@url [1]{\endgroup\@href {#1}{\urlprefix }}%
\providecommand \urlprefix  [0]{URL }%
\providecommand \Eprint [0]{\href }%
\providecommand \doibase [0]{https://doi.org/}%
\providecommand \selectlanguage [0]{\@gobble}%
\providecommand \bibinfo  [0]{\@secondoftwo}%
\providecommand \bibfield  [0]{\@secondoftwo}%
\providecommand \translation [1]{[#1]}%
\providecommand \BibitemOpen [0]{}%
\providecommand \bibitemStop [0]{}%
\providecommand \bibitemNoStop [0]{.\EOS\space}%
\providecommand \EOS [0]{\spacefactor3000\relax}%
\providecommand \BibitemShut  [1]{\csname bibitem#1\endcsname}%
\let\auto@bib@innerbib\@empty
\bibitem [{\citenamefont {Andrei}\ \emph {et~al.}(2021)\citenamefont {Andrei}, \citenamefont {Efetov}, \citenamefont {Jarillo-Herrero}, \citenamefont {MacDonald}, \citenamefont {Mak}, \citenamefont {Senthil}, \citenamefont {Tutuc}, \citenamefont {Yazdani},\ and\ \citenamefont {Young}}]{Andrei2021}%
  \BibitemOpen
  \bibfield  {author} {\bibinfo {author} {\bibfnamefont {E.~Y.}\ \bibnamefont {Andrei}}, \bibinfo {author} {\bibfnamefont {D.~K.}\ \bibnamefont {Efetov}}, \bibinfo {author} {\bibfnamefont {P.}~\bibnamefont {Jarillo-Herrero}}, \bibinfo {author} {\bibfnamefont {A.~H.}\ \bibnamefont {MacDonald}}, \bibinfo {author} {\bibfnamefont {K.~F.}\ \bibnamefont {Mak}}, \bibinfo {author} {\bibfnamefont {T.}~\bibnamefont {Senthil}}, \bibinfo {author} {\bibfnamefont {E.}~\bibnamefont {Tutuc}}, \bibinfo {author} {\bibfnamefont {A.}~\bibnamefont {Yazdani}},\ and\ \bibinfo {author} {\bibfnamefont {A.~F.}\ \bibnamefont {Young}},\ }\bibfield  {title} {\bibinfo {title} {The marvels of moir\'e materials},\ }\href {https://doi.org/10.1038/s41578-021-00284-1} {\bibfield  {journal} {\bibinfo  {journal} {Nat. Rev. Mater.}\ }\textbf {\bibinfo {volume} {6}},\ \bibinfo {pages} {201} (\bibinfo {year} {2021})}\BibitemShut {NoStop}%
\bibitem [{\citenamefont {Kennes}\ \emph {et~al.}(2021)\citenamefont {Kennes}, \citenamefont {Claassen}, \citenamefont {Xian}, \citenamefont {Georges}, \citenamefont {Millis}, \citenamefont {Hone}, \citenamefont {Dean}, \citenamefont {Basov},\ and\ \citenamefont {Pasupathy}}]{Kennes2021}%
  \BibitemOpen
  \bibfield  {author} {\bibinfo {author} {\bibfnamefont {D.~M.}\ \bibnamefont {Kennes}}, \bibinfo {author} {\bibfnamefont {M.}~\bibnamefont {Claassen}}, \bibinfo {author} {\bibfnamefont {L.}~\bibnamefont {Xian}}, \bibinfo {author} {\bibfnamefont {A.}~\bibnamefont {Georges}}, \bibinfo {author} {\bibfnamefont {A.~J.}\ \bibnamefont {Millis}}, \bibinfo {author} {\bibfnamefont {J.}~\bibnamefont {Hone}}, \bibinfo {author} {\bibfnamefont {C.~R.}\ \bibnamefont {Dean}}, \bibinfo {author} {\bibfnamefont {D.~N.}\ \bibnamefont {Basov}},\ and\ \bibinfo {author} {\bibfnamefont {A.~N.}\ \bibnamefont {Pasupathy}},\ }\bibfield  {title} {\bibinfo {title} {Moir\'e heterostructures as a condensed-matter quantum simulator},\ }\href {https://doi.org/10.1038/s41567-020-01154-3} {\bibfield  {journal} {\bibinfo  {journal} {Nat. Phys.}\ }\textbf {\bibinfo {volume} {17}},\ \bibinfo {pages} {155} (\bibinfo {year} {2021})}\BibitemShut {NoStop}%
\bibitem [{\citenamefont {Balents}\ \emph {et~al.}(2020)\citenamefont {Balents}, \citenamefont {Dean}, \citenamefont {Efetov},\ and\ \citenamefont {Young}}]{balents2020}%
  \BibitemOpen
  \bibfield  {author} {\bibinfo {author} {\bibfnamefont {L.}~\bibnamefont {Balents}}, \bibinfo {author} {\bibfnamefont {C.~R.}\ \bibnamefont {Dean}}, \bibinfo {author} {\bibfnamefont {D.~K.}\ \bibnamefont {Efetov}},\ and\ \bibinfo {author} {\bibfnamefont {A.~F.}\ \bibnamefont {Young}},\ }\bibfield  {title} {\bibinfo {title} {Superconductivity and strong correlations in moir\'e flat bands},\ }\href {https://doi.org/10.1038/s41567-020-0906-9} {\bibfield  {journal} {\bibinfo  {journal} {Nat. Phys.}\ }\textbf {\bibinfo {volume} {16}},\ \bibinfo {pages} {725} (\bibinfo {year} {2020})}\BibitemShut {NoStop}%
\bibitem [{\citenamefont {Carr}\ \emph {et~al.}(2020)\citenamefont {Carr}, \citenamefont {Fang},\ and\ \citenamefont {Kaxiras}}]{carr2020}%
  \BibitemOpen
  \bibfield  {author} {\bibinfo {author} {\bibfnamefont {S.}~\bibnamefont {Carr}}, \bibinfo {author} {\bibfnamefont {S.}~\bibnamefont {Fang}},\ and\ \bibinfo {author} {\bibfnamefont {E.}~\bibnamefont {Kaxiras}},\ }\bibfield  {title} {\bibinfo {title} {Electronic-structure methods for twisted moir\'e layers},\ }\href {https://doi.org/10.1038/s41578-020-0214-0} {\bibfield  {journal} {\bibinfo  {journal} {Nat. Rev. Mater.}\ }\textbf {\bibinfo {volume} {5}},\ \bibinfo {pages} {748} (\bibinfo {year} {2020})}\BibitemShut {NoStop}%
\bibitem [{\citenamefont {Mak}\ and\ \citenamefont {Shan}(2022)}]{Mak2022}%
  \BibitemOpen
  \bibfield  {author} {\bibinfo {author} {\bibfnamefont {K.~F.}\ \bibnamefont {Mak}}\ and\ \bibinfo {author} {\bibfnamefont {J.}~\bibnamefont {Shan}},\ }\bibfield  {title} {\bibinfo {title} {Moir\'e heterostructures as a condensed-matter quantum simulator},\ }\href {https://doi.org/10.1038/s41565-022-01165-6} {\bibfield  {journal} {\bibinfo  {journal} {Nat. Nanotechnol.}\ }\textbf {\bibinfo {volume} {17}},\ \bibinfo {pages} {686} (\bibinfo {year} {2022})}\BibitemShut {NoStop}%
\bibitem [{\citenamefont {Cai}\ \emph {et~al.}(2023)\citenamefont {Cai}, \citenamefont {Anderson}, \citenamefont {Wang}, \citenamefont {Zhang}, \citenamefont {Liu}, \citenamefont {Holtzmann}, \citenamefont {Zhang}, \citenamefont {Fan}, \citenamefont {Taniguchi}, \citenamefont {Watanabe}, \citenamefont {Ran}, \citenamefont {Cao}, \citenamefont {Fu}, \citenamefont {Xiao}, \citenamefont {Yao},\ and\ \citenamefont {Xu}}]{Cai2023}%
  \BibitemOpen
  \bibfield  {author} {\bibinfo {author} {\bibfnamefont {J.}~\bibnamefont {Cai}}, \bibinfo {author} {\bibfnamefont {E.}~\bibnamefont {Anderson}}, \bibinfo {author} {\bibfnamefont {C.}~\bibnamefont {Wang}}, \bibinfo {author} {\bibfnamefont {X.}~\bibnamefont {Zhang}}, \bibinfo {author} {\bibfnamefont {X.}~\bibnamefont {Liu}}, \bibinfo {author} {\bibfnamefont {W.}~\bibnamefont {Holtzmann}}, \bibinfo {author} {\bibfnamefont {Y.}~\bibnamefont {Zhang}}, \bibinfo {author} {\bibfnamefont {F.}~\bibnamefont {Fan}}, \bibinfo {author} {\bibfnamefont {T.}~\bibnamefont {Taniguchi}}, \bibinfo {author} {\bibfnamefont {K.}~\bibnamefont {Watanabe}}, \bibinfo {author} {\bibfnamefont {Y.}~\bibnamefont {Ran}}, \bibinfo {author} {\bibfnamefont {T.}~\bibnamefont {Cao}}, \bibinfo {author} {\bibfnamefont {L.}~\bibnamefont {Fu}}, \bibinfo {author} {\bibfnamefont {D.}~\bibnamefont {Xiao}}, \bibinfo {author} {\bibfnamefont {W.}~\bibnamefont {Yao}},\ and\ \bibinfo {author} {\bibfnamefont {X.}~\bibnamefont {Xu}},\ }\bibfield  {title}
  {\bibinfo {title} {Signatures of fractional quantum anomalous hall states in twisted {M}o{T}e$_{2}$},\ }\href {https://doi.org/10.1038/s41586-023-06289-w} {\bibfield  {journal} {\bibinfo  {journal} {Nature}\ }\textbf {\bibinfo {volume} {622}},\ \bibinfo {pages} {63} (\bibinfo {year} {2023})}\BibitemShut {NoStop}%
\bibitem [{\citenamefont {Zeng}\ \emph {et~al.}(2023)\citenamefont {Zeng}, \citenamefont {Xia}, \citenamefont {Kang}, \citenamefont {Zhu}, \citenamefont {Knüppel}, \citenamefont {Vaswani}, \citenamefont {Watanabe}, \citenamefont {Taniguchi}, \citenamefont {Mak},\ and\ \citenamefont {Shan}}]{Zeng2023}%
  \BibitemOpen
  \bibfield  {author} {\bibinfo {author} {\bibfnamefont {Y.}~\bibnamefont {Zeng}}, \bibinfo {author} {\bibfnamefont {Z.}~\bibnamefont {Xia}}, \bibinfo {author} {\bibfnamefont {K.}~\bibnamefont {Kang}}, \bibinfo {author} {\bibfnamefont {J.}~\bibnamefont {Zhu}}, \bibinfo {author} {\bibfnamefont {P.}~\bibnamefont {Knüppel}}, \bibinfo {author} {\bibfnamefont {C.}~\bibnamefont {Vaswani}}, \bibinfo {author} {\bibfnamefont {K.}~\bibnamefont {Watanabe}}, \bibinfo {author} {\bibfnamefont {T.}~\bibnamefont {Taniguchi}}, \bibinfo {author} {\bibfnamefont {K.~F.}\ \bibnamefont {Mak}},\ and\ \bibinfo {author} {\bibfnamefont {J.}~\bibnamefont {Shan}},\ }\bibfield  {title} {\bibinfo {title} {Thermodynamic evidence of fractional chern insulator in moir\'e {M}o{T}e$_{2}$},\ }\href {https://doi.org/10.1038/s41586-023-06452-3} {\bibfield  {journal} {\bibinfo  {journal} {Nature}\ }\textbf {\bibinfo {volume} {622}},\ \bibinfo {pages} {69} (\bibinfo {year} {2023})}\BibitemShut {NoStop}%
\bibitem [{\citenamefont {Park}\ \emph {et~al.}(2023)\citenamefont {Park}, \citenamefont {Cai}, \citenamefont {Anderson}, \citenamefont {Zhang}, \citenamefont {Zhu}, \citenamefont {Liu}, \citenamefont {Wang}, \citenamefont {Holtzmann}, \citenamefont {Hu}, \citenamefont {Liu}, \citenamefont {Taniguchi}, \citenamefont {Watanabe}, \citenamefont {Chu}, \citenamefont {Cao}, \citenamefont {Fu}, \citenamefont {Yao}, \citenamefont {Chang}, \citenamefont {Cobden}, \citenamefont {Xiao},\ and\ \citenamefont {Xu}}]{Park2023}%
  \BibitemOpen
  \bibfield  {author} {\bibinfo {author} {\bibfnamefont {H.}~\bibnamefont {Park}}, \bibinfo {author} {\bibfnamefont {J.}~\bibnamefont {Cai}}, \bibinfo {author} {\bibfnamefont {E.}~\bibnamefont {Anderson}}, \bibinfo {author} {\bibfnamefont {Y.}~\bibnamefont {Zhang}}, \bibinfo {author} {\bibfnamefont {J.}~\bibnamefont {Zhu}}, \bibinfo {author} {\bibfnamefont {X.}~\bibnamefont {Liu}}, \bibinfo {author} {\bibfnamefont {C.}~\bibnamefont {Wang}}, \bibinfo {author} {\bibfnamefont {W.}~\bibnamefont {Holtzmann}}, \bibinfo {author} {\bibfnamefont {C.}~\bibnamefont {Hu}}, \bibinfo {author} {\bibfnamefont {Z.}~\bibnamefont {Liu}}, \bibinfo {author} {\bibfnamefont {T.}~\bibnamefont {Taniguchi}}, \bibinfo {author} {\bibfnamefont {K.}~\bibnamefont {Watanabe}}, \bibinfo {author} {\bibfnamefont {J.-H.}\ \bibnamefont {Chu}}, \bibinfo {author} {\bibfnamefont {T.}~\bibnamefont {Cao}}, \bibinfo {author} {\bibfnamefont {L.}~\bibnamefont {Fu}}, \bibinfo {author} {\bibfnamefont {W.}~\bibnamefont {Yao}}, \bibinfo {author}
  {\bibfnamefont {C.-Z.}\ \bibnamefont {Chang}}, \bibinfo {author} {\bibfnamefont {D.}~\bibnamefont {Cobden}}, \bibinfo {author} {\bibfnamefont {D.}~\bibnamefont {Xiao}},\ and\ \bibinfo {author} {\bibfnamefont {X.}~\bibnamefont {Xu}},\ }\bibfield  {title} {\bibinfo {title} {Observation of fractionally quantized anomalous hall effect},\ }\href {https://doi.org/10.1038/s41586-023-06536-0} {\bibfield  {journal} {\bibinfo  {journal} {Nature}\ }\textbf {\bibinfo {volume} {622}},\ \bibinfo {pages} {74} (\bibinfo {year} {2023})}\BibitemShut {NoStop}%
\bibitem [{\citenamefont {Xu}\ \emph {et~al.}(2023)\citenamefont {Xu}, \citenamefont {Sun}, \citenamefont {Jia}, \citenamefont {Liu}, \citenamefont {Xu}, \citenamefont {Li}, \citenamefont {Gu}, \citenamefont {Watanabe}, \citenamefont {Taniguchi}, \citenamefont {Tong}, \citenamefont {Jia}, \citenamefont {Shi}, \citenamefont {Jiang}, \citenamefont {Zhang}, \citenamefont {Liu},\ and\ \citenamefont {Li}}]{Xu2023}%
  \BibitemOpen
  \bibfield  {author} {\bibinfo {author} {\bibfnamefont {F.}~\bibnamefont {Xu}}, \bibinfo {author} {\bibfnamefont {Z.}~\bibnamefont {Sun}}, \bibinfo {author} {\bibfnamefont {T.}~\bibnamefont {Jia}}, \bibinfo {author} {\bibfnamefont {C.}~\bibnamefont {Liu}}, \bibinfo {author} {\bibfnamefont {C.}~\bibnamefont {Xu}}, \bibinfo {author} {\bibfnamefont {C.}~\bibnamefont {Li}}, \bibinfo {author} {\bibfnamefont {Y.}~\bibnamefont {Gu}}, \bibinfo {author} {\bibfnamefont {K.}~\bibnamefont {Watanabe}}, \bibinfo {author} {\bibfnamefont {T.}~\bibnamefont {Taniguchi}}, \bibinfo {author} {\bibfnamefont {B.}~\bibnamefont {Tong}}, \bibinfo {author} {\bibfnamefont {J.}~\bibnamefont {Jia}}, \bibinfo {author} {\bibfnamefont {Z.}~\bibnamefont {Shi}}, \bibinfo {author} {\bibfnamefont {S.}~\bibnamefont {Jiang}}, \bibinfo {author} {\bibfnamefont {Y.}~\bibnamefont {Zhang}}, \bibinfo {author} {\bibfnamefont {X.}~\bibnamefont {Liu}},\ and\ \bibinfo {author} {\bibfnamefont {T.}~\bibnamefont {Li}},\ }\bibfield  {title} {\bibinfo {title}
  {Observation of integer and fractional quantum anomalous hall effects in twisted bilayer {M}o{T}e$_{2}$},\ }\href {https://doi.org/10.1103/PhysRevX.13.031037} {\bibfield  {journal} {\bibinfo  {journal} {Phys. Rev. X}\ }\textbf {\bibinfo {volume} {13}},\ \bibinfo {pages} {031037} (\bibinfo {year} {2023})}\BibitemShut {NoStop}%
\bibitem [{\citenamefont {Lu}\ \emph {et~al.}(2024{\natexlab{a}})\citenamefont {Lu}, \citenamefont {Han}, \citenamefont {Yao}, \citenamefont {Reddy}, \citenamefont {Yang}, \citenamefont {Seo}, \citenamefont {Watanabe}, \citenamefont {Taniguchi}, \citenamefont {Fu},\ and\ \citenamefont {Ju}}]{Lu2024}%
  \BibitemOpen
  \bibfield  {author} {\bibinfo {author} {\bibfnamefont {Z.}~\bibnamefont {Lu}}, \bibinfo {author} {\bibfnamefont {T.}~\bibnamefont {Han}}, \bibinfo {author} {\bibfnamefont {Y.}~\bibnamefont {Yao}}, \bibinfo {author} {\bibfnamefont {A.~P.}\ \bibnamefont {Reddy}}, \bibinfo {author} {\bibfnamefont {J.}~\bibnamefont {Yang}}, \bibinfo {author} {\bibfnamefont {J.}~\bibnamefont {Seo}}, \bibinfo {author} {\bibfnamefont {K.}~\bibnamefont {Watanabe}}, \bibinfo {author} {\bibfnamefont {T.}~\bibnamefont {Taniguchi}}, \bibinfo {author} {\bibfnamefont {L.}~\bibnamefont {Fu}},\ and\ \bibinfo {author} {\bibfnamefont {L.}~\bibnamefont {Ju}},\ }\bibfield  {title} {\bibinfo {title} {Fractional quantum anomalous hall effect in multilayer graphene},\ }\href {https://doi.org/10.1038/s41586-023-07010-7} {\bibfield  {journal} {\bibinfo  {journal} {Nature}\ }\textbf {\bibinfo {volume} {626}},\ \bibinfo {pages} {759} (\bibinfo {year} {2024}{\natexlab{a}})}\BibitemShut {NoStop}%
\bibitem [{\citenamefont {Li}\ \emph {et~al.}(2021)\citenamefont {Li}, \citenamefont {Kumar}, \citenamefont {Sun},\ and\ \citenamefont {Lin}}]{li2021}%
  \BibitemOpen
  \bibfield  {author} {\bibinfo {author} {\bibfnamefont {H.}~\bibnamefont {Li}}, \bibinfo {author} {\bibfnamefont {U.}~\bibnamefont {Kumar}}, \bibinfo {author} {\bibfnamefont {K.}~\bibnamefont {Sun}},\ and\ \bibinfo {author} {\bibfnamefont {S.-Z.}\ \bibnamefont {Lin}},\ }\bibfield  {title} {\bibinfo {title} {Spontaneous fractional chern insulators in transition metal dichalcogenide moir\'e superlattices},\ }\href {https://doi.org/10.1103/PhysRevResearch.3.L032070} {\bibfield  {journal} {\bibinfo  {journal} {Phys. Rev. Res.}\ }\textbf {\bibinfo {volume} {3}},\ \bibinfo {pages} {L032070} (\bibinfo {year} {2021})}\BibitemShut {NoStop}%
\bibitem [{\citenamefont {Devakul}\ \emph {et~al.}(2021)\citenamefont {Devakul}, \citenamefont {Cr{\'e}pel}, \citenamefont {Zhang},\ and\ \citenamefont {Fu}}]{devakul2021}%
  \BibitemOpen
  \bibfield  {author} {\bibinfo {author} {\bibfnamefont {T.}~\bibnamefont {Devakul}}, \bibinfo {author} {\bibfnamefont {V.}~\bibnamefont {Cr{\'e}pel}}, \bibinfo {author} {\bibfnamefont {Y.}~\bibnamefont {Zhang}},\ and\ \bibinfo {author} {\bibfnamefont {L.}~\bibnamefont {Fu}},\ }\bibfield  {title} {\bibinfo {title} {Magic in twisted transition metal dichalcogenide bilayers},\ }\href {https://doi.org/10.1038/s41467-021-27042-9} {\bibfield  {journal} {\bibinfo  {journal} {Nat. Commun.}\ }\textbf {\bibinfo {volume} {12}},\ \bibinfo {pages} {6730} (\bibinfo {year} {2021})}\BibitemShut {NoStop}%
\bibitem [{\citenamefont {Wang}\ \emph {et~al.}(2024)\citenamefont {Wang}, \citenamefont {Zhang}, \citenamefont {Liu}, \citenamefont {He}, \citenamefont {Xu}, \citenamefont {Ran}, \citenamefont {Cao},\ and\ \citenamefont {Xiao}}]{Wang2024}%
  \BibitemOpen
  \bibfield  {author} {\bibinfo {author} {\bibfnamefont {C.}~\bibnamefont {Wang}}, \bibinfo {author} {\bibfnamefont {X.-W.}\ \bibnamefont {Zhang}}, \bibinfo {author} {\bibfnamefont {X.}~\bibnamefont {Liu}}, \bibinfo {author} {\bibfnamefont {Y.}~\bibnamefont {He}}, \bibinfo {author} {\bibfnamefont {X.}~\bibnamefont {Xu}}, \bibinfo {author} {\bibfnamefont {Y.}~\bibnamefont {Ran}}, \bibinfo {author} {\bibfnamefont {T.}~\bibnamefont {Cao}},\ and\ \bibinfo {author} {\bibfnamefont {D.}~\bibnamefont {Xiao}},\ }\bibfield  {title} {\bibinfo {title} {Fractional chern insulator in twisted bilayer {M}o{T}e$_{2}$},\ }\href {https://doi.org/10.1103/PhysRevLett.132.036501} {\bibfield  {journal} {\bibinfo  {journal} {Phys. Rev. Lett.}\ }\textbf {\bibinfo {volume} {132}},\ \bibinfo {pages} {036501} (\bibinfo {year} {2024})}\BibitemShut {NoStop}%
\bibitem [{\citenamefont {Jia}\ \emph {et~al.}(2024)\citenamefont {Jia}, \citenamefont {Yu}, \citenamefont {Liu}, \citenamefont {Herzog-Arbeitman}, \citenamefont {Qi}, \citenamefont {Pi}, \citenamefont {Regnault}, \citenamefont {Weng}, \citenamefont {Bernevig},\ and\ \citenamefont {Wu}}]{Jia2024}%
  \BibitemOpen
  \bibfield  {author} {\bibinfo {author} {\bibfnamefont {Y.}~\bibnamefont {Jia}}, \bibinfo {author} {\bibfnamefont {J.}~\bibnamefont {Yu}}, \bibinfo {author} {\bibfnamefont {J.}~\bibnamefont {Liu}}, \bibinfo {author} {\bibfnamefont {J.}~\bibnamefont {Herzog-Arbeitman}}, \bibinfo {author} {\bibfnamefont {Z.}~\bibnamefont {Qi}}, \bibinfo {author} {\bibfnamefont {H.}~\bibnamefont {Pi}}, \bibinfo {author} {\bibfnamefont {N.}~\bibnamefont {Regnault}}, \bibinfo {author} {\bibfnamefont {H.}~\bibnamefont {Weng}}, \bibinfo {author} {\bibfnamefont {B.~A.}\ \bibnamefont {Bernevig}},\ and\ \bibinfo {author} {\bibfnamefont {Q.}~\bibnamefont {Wu}},\ }\bibfield  {title} {\bibinfo {title} {Moir\'e fractional chern insulators. i. first-principles calculations and continuum models of twisted bilayer {M}o{T}e$_{2}$},\ }\href {https://doi.org/10.1103/PhysRevB.109.205121} {\bibfield  {journal} {\bibinfo  {journal} {Phys. Rev. B}\ }\textbf {\bibinfo {volume} {109}},\ \bibinfo {pages} {205121} (\bibinfo {year} {2024})}\BibitemShut
  {NoStop}%
\bibitem [{\citenamefont {Yu}\ \emph {et~al.}(2024)\citenamefont {Yu}, \citenamefont {Herzog-Arbeitman}, \citenamefont {Wang}, \citenamefont {Vafek}, \citenamefont {Bernevig},\ and\ \citenamefont {Regnault}}]{Yu2024}%
  \BibitemOpen
  \bibfield  {author} {\bibinfo {author} {\bibfnamefont {J.}~\bibnamefont {Yu}}, \bibinfo {author} {\bibfnamefont {J.}~\bibnamefont {Herzog-Arbeitman}}, \bibinfo {author} {\bibfnamefont {M.}~\bibnamefont {Wang}}, \bibinfo {author} {\bibfnamefont {O.}~\bibnamefont {Vafek}}, \bibinfo {author} {\bibfnamefont {B.~A.}\ \bibnamefont {Bernevig}},\ and\ \bibinfo {author} {\bibfnamefont {N.}~\bibnamefont {Regnault}},\ }\bibfield  {title} {\bibinfo {title} {Fractional chern insulators versus nonmagnetic states in twisted bilayer {M}o{T}e$_{2}$},\ }\href {https://doi.org/10.1103/PhysRevB.109.045147} {\bibfield  {journal} {\bibinfo  {journal} {Phys. Rev. B}\ }\textbf {\bibinfo {volume} {109}},\ \bibinfo {pages} {045147} (\bibinfo {year} {2024})}\BibitemShut {NoStop}%
\bibitem [{\citenamefont {Morales-Dur\'an}\ \emph {et~al.}(2024)\citenamefont {Morales-Dur\'an}, \citenamefont {Wei}, \citenamefont {Shi},\ and\ \citenamefont {MacDonald}}]{macdonald2024}%
  \BibitemOpen
  \bibfield  {author} {\bibinfo {author} {\bibfnamefont {N.}~\bibnamefont {Morales-Dur\'an}}, \bibinfo {author} {\bibfnamefont {N.}~\bibnamefont {Wei}}, \bibinfo {author} {\bibfnamefont {J.}~\bibnamefont {Shi}},\ and\ \bibinfo {author} {\bibfnamefont {A.~H.}\ \bibnamefont {MacDonald}},\ }\bibfield  {title} {\bibinfo {title} {Magic angles and fractional chern insulators in twisted homobilayer transition metal dichalcogenides},\ }\href {https://doi.org/10.1103/PhysRevLett.132.096602} {\bibfield  {journal} {\bibinfo  {journal} {Phys. Rev. Lett.}\ }\textbf {\bibinfo {volume} {132}},\ \bibinfo {pages} {096602} (\bibinfo {year} {2024})}\BibitemShut {NoStop}%
\bibitem [{\citenamefont {Cao}\ \emph {et~al.}(2018)\citenamefont {Cao}, \citenamefont {Fatemi}, \citenamefont {Fang}, \citenamefont {Watanabe}, \citenamefont {Taniguchi}, \citenamefont {Kaxiras},\ and\ \citenamefont {Jarillo-Herrero}}]{Cao2018_2}%
  \BibitemOpen
  \bibfield  {author} {\bibinfo {author} {\bibfnamefont {Y.}~\bibnamefont {Cao}}, \bibinfo {author} {\bibfnamefont {V.}~\bibnamefont {Fatemi}}, \bibinfo {author} {\bibfnamefont {S.}~\bibnamefont {Fang}}, \bibinfo {author} {\bibfnamefont {K.}~\bibnamefont {Watanabe}}, \bibinfo {author} {\bibfnamefont {T.}~\bibnamefont {Taniguchi}}, \bibinfo {author} {\bibfnamefont {E.}~\bibnamefont {Kaxiras}},\ and\ \bibinfo {author} {\bibfnamefont {P.}~\bibnamefont {Jarillo-Herrero}},\ }\bibfield  {title} {\bibinfo {title} {Unconventional superconductivity in magic-angle graphene superlattices},\ }\href {https://doi.org/10.1038/nature26160} {\bibfield  {journal} {\bibinfo  {journal} {Nature}\ }\textbf {\bibinfo {volume} {556}},\ \bibinfo {pages} {43} (\bibinfo {year} {2018})}\BibitemShut {NoStop}%
\bibitem [{\citenamefont {Yankowitz}\ \emph {et~al.}(2019)\citenamefont {Yankowitz}, \citenamefont {Chen}, \citenamefont {Polshyn}, \citenamefont {Zhang}, \citenamefont {Watanabe}, \citenamefont {Taniguchi}, \citenamefont {Graf}, \citenamefont {Young},\ and\ \citenamefont {Dean}}]{yankowitz2019}%
  \BibitemOpen
  \bibfield  {author} {\bibinfo {author} {\bibfnamefont {M.}~\bibnamefont {Yankowitz}}, \bibinfo {author} {\bibfnamefont {S.}~\bibnamefont {Chen}}, \bibinfo {author} {\bibfnamefont {H.}~\bibnamefont {Polshyn}}, \bibinfo {author} {\bibfnamefont {Y.}~\bibnamefont {Zhang}}, \bibinfo {author} {\bibfnamefont {K.}~\bibnamefont {Watanabe}}, \bibinfo {author} {\bibfnamefont {T.}~\bibnamefont {Taniguchi}}, \bibinfo {author} {\bibfnamefont {D.}~\bibnamefont {Graf}}, \bibinfo {author} {\bibfnamefont {A.~F.}\ \bibnamefont {Young}},\ and\ \bibinfo {author} {\bibfnamefont {C.~R.}\ \bibnamefont {Dean}},\ }\bibfield  {title} {\bibinfo {title} {Tuning superconductivity in twisted bilayer graphene},\ }\href {https://doi.org/10.1126/science.aav1910} {\bibfield  {journal} {\bibinfo  {journal} {Science}\ }\textbf {\bibinfo {volume} {363}},\ \bibinfo {pages} {1059} (\bibinfo {year} {2019})}\BibitemShut {NoStop}%
\bibitem [{\citenamefont {Xia}\ \emph {et~al.}(2025)\citenamefont {Xia}, \citenamefont {Han}, \citenamefont {Watanabe}, \citenamefont {Taniguchi}, \citenamefont {Shan},\ and\ \citenamefont {Mak}}]{Xia2024}%
  \BibitemOpen
  \bibfield  {author} {\bibinfo {author} {\bibfnamefont {Y.}~\bibnamefont {Xia}}, \bibinfo {author} {\bibfnamefont {Z.}~\bibnamefont {Han}}, \bibinfo {author} {\bibfnamefont {K.}~\bibnamefont {Watanabe}}, \bibinfo {author} {\bibfnamefont {T.}~\bibnamefont {Taniguchi}}, \bibinfo {author} {\bibfnamefont {J.}~\bibnamefont {Shan}},\ and\ \bibinfo {author} {\bibfnamefont {K.~F.}\ \bibnamefont {Mak}},\ }\bibfield  {title} {\bibinfo {title} {Superconductivity in twisted bilayer {WS}e$_{2}$},\ }\href {https://doi.org/10.1038/s41586-024-08116-2} {\bibfield  {journal} {\bibinfo  {journal} {Nature}\ }\textbf {\bibinfo {volume} {637}},\ \bibinfo {pages} {833} (\bibinfo {year} {2025})}\BibitemShut {NoStop}%
\bibitem [{\citenamefont {Guo}\ \emph {et~al.}(2025)\citenamefont {Guo}, \citenamefont {Pack}, \citenamefont {Swann}, \citenamefont {Holtzman}, \citenamefont {Cothrine}, \citenamefont {Watanabe}, \citenamefont {Taniguchi}, \citenamefont {Mandrus}, \citenamefont {Barmak}, \citenamefont {Hone} \emph {et~al.}}]{Guo2025}%
  \BibitemOpen
  \bibfield  {author} {\bibinfo {author} {\bibfnamefont {Y.}~\bibnamefont {Guo}}, \bibinfo {author} {\bibfnamefont {J.}~\bibnamefont {Pack}}, \bibinfo {author} {\bibfnamefont {J.}~\bibnamefont {Swann}}, \bibinfo {author} {\bibfnamefont {L.}~\bibnamefont {Holtzman}}, \bibinfo {author} {\bibfnamefont {M.}~\bibnamefont {Cothrine}}, \bibinfo {author} {\bibfnamefont {K.}~\bibnamefont {Watanabe}}, \bibinfo {author} {\bibfnamefont {T.}~\bibnamefont {Taniguchi}}, \bibinfo {author} {\bibfnamefont {D.~G.}\ \bibnamefont {Mandrus}}, \bibinfo {author} {\bibfnamefont {K.}~\bibnamefont {Barmak}}, \bibinfo {author} {\bibfnamefont {J.}~\bibnamefont {Hone}}, \emph {et~al.},\ }\bibfield  {title} {\bibinfo {title} {Superconductivity in 5.0$^\circ$ twisted bilayer {WS}e$_{2}$},\ }\href {https://doi.org/10.1038/s41586-024-08381-1} {\bibfield  {journal} {\bibinfo  {journal} {Nature}\ }\textbf {\bibinfo {volume} {637}},\ \bibinfo {pages} {839} (\bibinfo {year} {2025})}\BibitemShut {NoStop}%
\bibitem [{\citenamefont {Kn{\"u}ppel}\ \emph {et~al.}(2025)\citenamefont {Kn{\"u}ppel}, \citenamefont {Zhu}, \citenamefont {Xia}, \citenamefont {Xia}, \citenamefont {Han}, \citenamefont {Zeng}, \citenamefont {Watanabe}, \citenamefont {Taniguchi}, \citenamefont {Shan},\ and\ \citenamefont {Mak}}]{knuppel2025correlated}%
  \BibitemOpen
  \bibfield  {author} {\bibinfo {author} {\bibfnamefont {P.}~\bibnamefont {Kn{\"u}ppel}}, \bibinfo {author} {\bibfnamefont {J.}~\bibnamefont {Zhu}}, \bibinfo {author} {\bibfnamefont {Y.}~\bibnamefont {Xia}}, \bibinfo {author} {\bibfnamefont {Z.}~\bibnamefont {Xia}}, \bibinfo {author} {\bibfnamefont {Z.}~\bibnamefont {Han}}, \bibinfo {author} {\bibfnamefont {Y.}~\bibnamefont {Zeng}}, \bibinfo {author} {\bibfnamefont {K.}~\bibnamefont {Watanabe}}, \bibinfo {author} {\bibfnamefont {T.}~\bibnamefont {Taniguchi}}, \bibinfo {author} {\bibfnamefont {J.}~\bibnamefont {Shan}},\ and\ \bibinfo {author} {\bibfnamefont {K.~F.}\ \bibnamefont {Mak}},\ }\bibfield  {title} {\bibinfo {title} {Correlated states controlled by a tunable van hove singularity in moir\'e {WS}e$_2$ bilayers},\ }\href {https://doi.org/https://doi.org/10.1038/s41467-025-57235-5} {\bibfield  {journal} {\bibinfo  {journal} {Nat. Commun.}\ }\textbf {\bibinfo {volume} {16}},\ \bibinfo {pages} {1959} (\bibinfo {year} {2025})}\BibitemShut {NoStop}%
\bibitem [{\citenamefont {Wu}\ \emph {et~al.}(2018)\citenamefont {Wu}, \citenamefont {Lovorn}, \citenamefont {Tutuc},\ and\ \citenamefont {MacDonald}}]{Wu2018}%
  \BibitemOpen
  \bibfield  {author} {\bibinfo {author} {\bibfnamefont {F.}~\bibnamefont {Wu}}, \bibinfo {author} {\bibfnamefont {T.}~\bibnamefont {Lovorn}}, \bibinfo {author} {\bibfnamefont {E.}~\bibnamefont {Tutuc}},\ and\ \bibinfo {author} {\bibfnamefont {A.~H.}\ \bibnamefont {MacDonald}},\ }\bibfield  {title} {\bibinfo {title} {Hubbard model physics in transition metal dichalcogenide moir\'e bands},\ }\href {https://doi.org/10.1103/PhysRevLett.121.026402} {\bibfield  {journal} {\bibinfo  {journal} {Phys. Rev. Lett.}\ }\textbf {\bibinfo {volume} {121}},\ \bibinfo {pages} {026402} (\bibinfo {year} {2018})}\BibitemShut {NoStop}%
\bibitem [{\citenamefont {Wu}\ \emph {et~al.}(2019)\citenamefont {Wu}, \citenamefont {Lovorn}, \citenamefont {Tutuc}, \citenamefont {Martin},\ and\ \citenamefont {MacDonald}}]{Wu2019_tmd}%
  \BibitemOpen
  \bibfield  {author} {\bibinfo {author} {\bibfnamefont {F.}~\bibnamefont {Wu}}, \bibinfo {author} {\bibfnamefont {T.}~\bibnamefont {Lovorn}}, \bibinfo {author} {\bibfnamefont {E.}~\bibnamefont {Tutuc}}, \bibinfo {author} {\bibfnamefont {I.}~\bibnamefont {Martin}},\ and\ \bibinfo {author} {\bibfnamefont {A.~H.}\ \bibnamefont {MacDonald}},\ }\bibfield  {title} {\bibinfo {title} {Topological insulators in twisted transition metal dichalcogenide homobilayers},\ }\href {https://doi.org/10.1103/PhysRevLett.122.086402} {\bibfield  {journal} {\bibinfo  {journal} {Phys. Rev. Lett.}\ }\textbf {\bibinfo {volume} {122}},\ \bibinfo {pages} {086402} (\bibinfo {year} {2019})}\BibitemShut {NoStop}%
\bibitem [{\citenamefont {Bistritzer}\ and\ \citenamefont {MacDonald}(2011)}]{Bistritzer2011}%
  \BibitemOpen
  \bibfield  {author} {\bibinfo {author} {\bibfnamefont {R.}~\bibnamefont {Bistritzer}}\ and\ \bibinfo {author} {\bibfnamefont {A.~H.}\ \bibnamefont {MacDonald}},\ }\bibfield  {title} {\bibinfo {title} {Moir\'e bands in twisted double-layer graphene},\ }\href {https://doi.org/10.1073/pnas.1108174108} {\bibfield  {journal} {\bibinfo  {journal} {Proc. Natl. Acad. Sci. U.S.A.}\ }\textbf {\bibinfo {volume} {108}},\ \bibinfo {pages} {12233} (\bibinfo {year} {2011})}\BibitemShut {NoStop}%
\bibitem [{\citenamefont {Angeli}\ and\ \citenamefont {MacDonald}(2021)}]{Angeli2021}%
  \BibitemOpen
  \bibfield  {author} {\bibinfo {author} {\bibfnamefont {M.}~\bibnamefont {Angeli}}\ and\ \bibinfo {author} {\bibfnamefont {A.~H.}\ \bibnamefont {MacDonald}},\ }\bibfield  {title} {\bibinfo {title} {$\mathrm{\Gamma}$ valley transition metal dichalcogenide moir\'e bands},\ }\href {https://doi.org/10.1073/pnas.2021826118} {\bibfield  {journal} {\bibinfo  {journal} {Proc. Natl. Acad. Sci. U.S.A.}\ }\textbf {\bibinfo {volume} {118}},\ \bibinfo {pages} {e2021826118} (\bibinfo {year} {2021})}\BibitemShut {NoStop}%
\bibitem [{\citenamefont {Călugăru}\ \emph {et~al.}(2025)\citenamefont {Călugăru}, \citenamefont {Jiang}, \citenamefont {Hu}, \citenamefont {Pi}, \citenamefont {Yu}, \citenamefont {Vergniory}, \citenamefont {Shan}, \citenamefont {Felser}, \citenamefont {Schoop}, \citenamefont {Efetov}, \citenamefont {Mak},\ and\ \citenamefont {Bernevig}}]{Calugaru2025}%
  \BibitemOpen
  \bibfield  {author} {\bibinfo {author} {\bibfnamefont {D.}~\bibnamefont {Călugăru}}, \bibinfo {author} {\bibfnamefont {Y.}~\bibnamefont {Jiang}}, \bibinfo {author} {\bibfnamefont {H.}~\bibnamefont {Hu}}, \bibinfo {author} {\bibfnamefont {H.}~\bibnamefont {Pi}}, \bibinfo {author} {\bibfnamefont {J.}~\bibnamefont {Yu}}, \bibinfo {author} {\bibfnamefont {M.~G.}\ \bibnamefont {Vergniory}}, \bibinfo {author} {\bibfnamefont {J.}~\bibnamefont {Shan}}, \bibinfo {author} {\bibfnamefont {C.}~\bibnamefont {Felser}}, \bibinfo {author} {\bibfnamefont {L.~M.}\ \bibnamefont {Schoop}}, \bibinfo {author} {\bibfnamefont {D.~K.}\ \bibnamefont {Efetov}}, \bibinfo {author} {\bibfnamefont {K.~F.}\ \bibnamefont {Mak}},\ and\ \bibinfo {author} {\bibfnamefont {B.~A.}\ \bibnamefont {Bernevig}},\ }\bibfield  {title} {\bibinfo {title} {Moiré materials based on m-point twisting},\ }\href {https://doi.org/10.1038/s41586-025-09187-5} {\bibfield  {journal} {\bibinfo  {journal} {Nature}\ }\textbf {\bibinfo {volume} {643}},\ \bibinfo
  {pages} {376} (\bibinfo {year} {2025})}\BibitemShut {NoStop}%
\bibitem [{\citenamefont {Bao}\ \emph {et~al.}(2025)\citenamefont {Bao}, \citenamefont {Wang}, \citenamefont {Liu},\ and\ \citenamefont {Wang}}]{Bao2025}%
  \BibitemOpen
  \bibfield  {author} {\bibinfo {author} {\bibfnamefont {K.}~\bibnamefont {Bao}}, \bibinfo {author} {\bibfnamefont {H.}~\bibnamefont {Wang}}, \bibinfo {author} {\bibfnamefont {Z.}~\bibnamefont {Liu}},\ and\ \bibinfo {author} {\bibfnamefont {J.}~\bibnamefont {Wang}},\ }\bibfield  {title} {\bibinfo {title} {Anisotropic moir\'e band flattening in twisted bilayers of {M}-valley mxenes},\ }\href {https://doi.org/10.1103/l88b-67d2} {\bibfield  {journal} {\bibinfo  {journal} {Phys. Rev. B}\ }\textbf {\bibinfo {volume} {112}},\ \bibinfo {pages} {L041406} (\bibinfo {year} {2025})}\BibitemShut {NoStop}%
\bibitem [{\citenamefont {Lei}\ \emph {et~al.}(2025)\citenamefont {Lei}, \citenamefont {Mahon},\ and\ \citenamefont {MacDonald}}]{Lei2025}%
  \BibitemOpen
  \bibfield  {author} {\bibinfo {author} {\bibfnamefont {C.}~\bibnamefont {Lei}}, \bibinfo {author} {\bibfnamefont {P.~T.}\ \bibnamefont {Mahon}},\ and\ \bibinfo {author} {\bibfnamefont {A.~H.}\ \bibnamefont {MacDonald}},\ }\bibfield  {title} {\bibinfo {title} {Moir\'e band theory for m-valley twisted transition metal dichalcogenides},\ }\href {https://doi.org/10.1103/5zt2-scbg} {\bibfield  {journal} {\bibinfo  {journal} {Phys. Rev. Lett.}\ }\textbf {\bibinfo {volume} {135}},\ \bibinfo {pages} {196402} (\bibinfo {year} {2025})}\BibitemShut {NoStop}%
\bibitem [{\citenamefont {Kariyado}\ and\ \citenamefont {Vishwanath}(2019)}]{kariyado2019}%
  \BibitemOpen
  \bibfield  {author} {\bibinfo {author} {\bibfnamefont {T.}~\bibnamefont {Kariyado}}\ and\ \bibinfo {author} {\bibfnamefont {A.}~\bibnamefont {Vishwanath}},\ }\bibfield  {title} {\bibinfo {title} {Flat band in twisted bilayer bravais lattices},\ }\href {https://doi.org/10.1103/PhysRevResearch.1.033076} {\bibfield  {journal} {\bibinfo  {journal} {Phys. Rev. Res.}\ }\textbf {\bibinfo {volume} {1}},\ \bibinfo {pages} {033076} (\bibinfo {year} {2019})}\BibitemShut {NoStop}%
\bibitem [{\citenamefont {Luo}\ \emph {et~al.}(2021)\citenamefont {Luo}, \citenamefont {Xu},\ and\ \citenamefont {Jian}}]{luo2021}%
  \BibitemOpen
  \bibfield  {author} {\bibinfo {author} {\bibfnamefont {Z.-X.}\ \bibnamefont {Luo}}, \bibinfo {author} {\bibfnamefont {C.}~\bibnamefont {Xu}},\ and\ \bibinfo {author} {\bibfnamefont {C.-M.}\ \bibnamefont {Jian}},\ }\bibfield  {title} {\bibinfo {title} {Magic continuum in a twisted bilayer square lattice with staggered flux},\ }\href {https://doi.org/10.1103/PhysRevB.104.035136} {\bibfield  {journal} {\bibinfo  {journal} {Phys. Rev. B}\ }\textbf {\bibinfo {volume} {104}},\ \bibinfo {pages} {035136} (\bibinfo {year} {2021})}\BibitemShut {NoStop}%
\bibitem [{\citenamefont {Eugenio}\ \emph {et~al.}(2025)\citenamefont {Eugenio}, \citenamefont {Luo}, \citenamefont {Vishwanath},\ and\ \citenamefont {Volkov}}]{Eugenio2025}%
  \BibitemOpen
  \bibfield  {author} {\bibinfo {author} {\bibfnamefont {P.~M.}\ \bibnamefont {Eugenio}}, \bibinfo {author} {\bibfnamefont {Z.-X.}\ \bibnamefont {Luo}}, \bibinfo {author} {\bibfnamefont {A.}~\bibnamefont {Vishwanath}},\ and\ \bibinfo {author} {\bibfnamefont {P.~A.}\ \bibnamefont {Volkov}},\ }\bibfield  {title} {\bibinfo {title} {Tunable $t\text{\ensuremath{-}}\ensuremath{t}^{\ensuremath{'}}\text{\ensuremath{-}}\ensuremath{U}$ hubbard models in twisted square homobilayers},\ }\href {https://doi.org/10.1103/PhysRevLett.134.236503} {\bibfield  {journal} {\bibinfo  {journal} {Phys. Rev. Lett.}\ }\textbf {\bibinfo {volume} {134}},\ \bibinfo {pages} {236503} (\bibinfo {year} {2025})}\BibitemShut {NoStop}%
\bibitem [{\citenamefont {Kariyado}\ \emph {et~al.}(2025)\citenamefont {Kariyado}, \citenamefont {Vishwanath},\ and\ \citenamefont {Luo}}]{Kariyado2025b}%
  \BibitemOpen
  \bibfield  {author} {\bibinfo {author} {\bibfnamefont {T.}~\bibnamefont {Kariyado}}, \bibinfo {author} {\bibfnamefont {A.}~\bibnamefont {Vishwanath}},\ and\ \bibinfo {author} {\bibfnamefont {Z.-X.}\ \bibnamefont {Luo}},\ }\bibfield  {title} {\bibinfo {title} {Single-band square-lattice hubbard model from twisted bilayer ${\mathrm{c}}_{568}$},\ }\href {https://doi.org/10.1103/73rf-cw9d} {\bibfield  {journal} {\bibinfo  {journal} {Phys. Rev. B}\ }\textbf {\bibinfo {volume} {112}},\ \bibinfo {pages} {125159} (\bibinfo {year} {2025})}\BibitemShut {NoStop}%
\bibitem [{\citenamefont {Xu}\ \emph {et~al.}(2025)\citenamefont {Xu}, \citenamefont {Fischer}, \citenamefont {Tancogne-Dejean}, \citenamefont {Zhang}, \citenamefont {Bostr\"om}, \citenamefont {Claassen}, \citenamefont {Kennes}, \citenamefont {Rubio},\ and\ \citenamefont {Xian}}]{xu2025engineering}%
  \BibitemOpen
  \bibfield  {author} {\bibinfo {author} {\bibfnamefont {Q.}~\bibnamefont {Xu}}, \bibinfo {author} {\bibfnamefont {A.}~\bibnamefont {Fischer}}, \bibinfo {author} {\bibfnamefont {N.}~\bibnamefont {Tancogne-Dejean}}, \bibinfo {author} {\bibfnamefont {T.}~\bibnamefont {Zhang}}, \bibinfo {author} {\bibfnamefont {E.~V.~n.}\ \bibnamefont {Bostr\"om}}, \bibinfo {author} {\bibfnamefont {M.}~\bibnamefont {Claassen}}, \bibinfo {author} {\bibfnamefont {D.~M.}\ \bibnamefont {Kennes}}, \bibinfo {author} {\bibfnamefont {A.}~\bibnamefont {Rubio}},\ and\ \bibinfo {author} {\bibfnamefont {L.}~\bibnamefont {Xian}},\ }\bibfield  {title} {\bibinfo {title} {Engineering 2d square lattice hubbard models in 90\ifmmode^\circ\else\textdegree\fi{} twisted $\mathrm{GeX}/\mathrm{SnX}$ ($\mathrm{X}=\mathrm{S}$, se) moir\'e superlattices},\ }\href {https://doi.org/10.1103/wcbz-lbr1} {\bibfield  {journal} {\bibinfo  {journal} {Phys. Rev. X}\ }\textbf {\bibinfo {volume} {15}},\ \bibinfo {pages} {041049} (\bibinfo {year} {2025})}\BibitemShut
  {NoStop}%
\bibitem [{\citenamefont {Soeda}\ \emph {et~al.}(2022)\citenamefont {Soeda}, \citenamefont {Asaga},\ and\ \citenamefont {Fukui}}]{soeda2022}%
  \BibitemOpen
  \bibfield  {author} {\bibinfo {author} {\bibfnamefont {Y.}~\bibnamefont {Soeda}}, \bibinfo {author} {\bibfnamefont {K.}~\bibnamefont {Asaga}},\ and\ \bibinfo {author} {\bibfnamefont {T.}~\bibnamefont {Fukui}},\ }\bibfield  {title} {\bibinfo {title} {Moir\'e landau levels of a ${C}_{4}$-symmetric twisted bilayer system in the absence of a magnetic field},\ }\href {https://doi.org/10.1103/PhysRevB.105.165422} {\bibfield  {journal} {\bibinfo  {journal} {Phys. Rev. B}\ }\textbf {\bibinfo {volume} {105}},\ \bibinfo {pages} {165422} (\bibinfo {year} {2022})}\BibitemShut {NoStop}%
\bibitem [{\citenamefont {Li}\ \emph {et~al.}(2022)\citenamefont {Li}, \citenamefont {He},\ and\ \citenamefont {Yao}}]{Li2022magic}%
  \BibitemOpen
  \bibfield  {author} {\bibinfo {author} {\bibfnamefont {M.-R.}\ \bibnamefont {Li}}, \bibinfo {author} {\bibfnamefont {A.-L.}\ \bibnamefont {He}},\ and\ \bibinfo {author} {\bibfnamefont {H.}~\bibnamefont {Yao}},\ }\bibfield  {title} {\bibinfo {title} {Magic-angle twisted bilayer systems with quadratic band touching: Exactly flat bands with high chern number},\ }\href {https://doi.org/10.1103/PhysRevResearch.4.043151} {\bibfield  {journal} {\bibinfo  {journal} {Phys. Rev. Res.}\ }\textbf {\bibinfo {volume} {4}},\ \bibinfo {pages} {043151} (\bibinfo {year} {2022})}\BibitemShut {NoStop}%
\bibitem [{\citenamefont {Sarkar}\ \emph {et~al.}(2025)\citenamefont {Sarkar}, \citenamefont {Wan}, \citenamefont {Lin},\ and\ \citenamefont {Sun}}]{sarkar2025}%
  \BibitemOpen
  \bibfield  {author} {\bibinfo {author} {\bibfnamefont {S.}~\bibnamefont {Sarkar}}, \bibinfo {author} {\bibfnamefont {X.}~\bibnamefont {Wan}}, \bibinfo {author} {\bibfnamefont {S.-Z.}\ \bibnamefont {Lin}},\ and\ \bibinfo {author} {\bibfnamefont {K.}~\bibnamefont {Sun}},\ }\bibfield  {title} {\bibinfo {title} {Symmetry-based classification of exact flat bands in single and bilayer moir\'e systems},\ }\href {https://doi.org/10.1103/nys8-5mg2} {\bibfield  {journal} {\bibinfo  {journal} {Phys. Rev. Lett.}\ }\textbf {\bibinfo {volume} {135}},\ \bibinfo {pages} {016501} (\bibinfo {year} {2025})}\BibitemShut {NoStop}%
\bibitem [{\citenamefont {Can}\ \emph {et~al.}(2021)\citenamefont {Can}, \citenamefont {Tummuru}, \citenamefont {Day}, \citenamefont {Elfimov}, \citenamefont {Damascelli},\ and\ \citenamefont {Franz}}]{can2021high}%
  \BibitemOpen
  \bibfield  {author} {\bibinfo {author} {\bibfnamefont {O.}~\bibnamefont {Can}}, \bibinfo {author} {\bibfnamefont {T.}~\bibnamefont {Tummuru}}, \bibinfo {author} {\bibfnamefont {R.~P.}\ \bibnamefont {Day}}, \bibinfo {author} {\bibfnamefont {I.}~\bibnamefont {Elfimov}}, \bibinfo {author} {\bibfnamefont {A.}~\bibnamefont {Damascelli}},\ and\ \bibinfo {author} {\bibfnamefont {M.}~\bibnamefont {Franz}},\ }\bibfield  {title} {\bibinfo {title} {High-temperature topological superconductivity in twisted double-layer copper oxides},\ }\href {https://doi.org/https://doi.org/10.1038/s41567-020-01142-7} {\bibfield  {journal} {\bibinfo  {journal} {Nature Physics}\ }\textbf {\bibinfo {volume} {17}},\ \bibinfo {pages} {519} (\bibinfo {year} {2021})}\BibitemShut {NoStop}%
\bibitem [{\citenamefont {Zhao}\ \emph {et~al.}(2023)\citenamefont {Zhao}, \citenamefont {Cui}, \citenamefont {Volkov}, \citenamefont {Yoo}, \citenamefont {Lee}, \citenamefont {Gardener}, \citenamefont {Akey}, \citenamefont {Engelke}, \citenamefont {Ronen}, \citenamefont {Zhong} \emph {et~al.}}]{zhao2023time}%
  \BibitemOpen
  \bibfield  {author} {\bibinfo {author} {\bibfnamefont {S.~F.}\ \bibnamefont {Zhao}}, \bibinfo {author} {\bibfnamefont {X.}~\bibnamefont {Cui}}, \bibinfo {author} {\bibfnamefont {P.~A.}\ \bibnamefont {Volkov}}, \bibinfo {author} {\bibfnamefont {H.}~\bibnamefont {Yoo}}, \bibinfo {author} {\bibfnamefont {S.}~\bibnamefont {Lee}}, \bibinfo {author} {\bibfnamefont {J.~A.}\ \bibnamefont {Gardener}}, \bibinfo {author} {\bibfnamefont {A.~J.}\ \bibnamefont {Akey}}, \bibinfo {author} {\bibfnamefont {R.}~\bibnamefont {Engelke}}, \bibinfo {author} {\bibfnamefont {Y.}~\bibnamefont {Ronen}}, \bibinfo {author} {\bibfnamefont {R.}~\bibnamefont {Zhong}}, \emph {et~al.},\ }\bibfield  {title} {\bibinfo {title} {Time-reversal symmetry breaking superconductivity between twisted cuprate superconductors},\ }\href {https://doi.org/10.1126/science.abl8371} {\bibfield  {journal} {\bibinfo  {journal} {Science}\ }\textbf {\bibinfo {volume} {382}},\ \bibinfo {pages} {1422} (\bibinfo {year} {2023})}\BibitemShut {NoStop}%
\bibitem [{\citenamefont {Song}\ \emph {et~al.}(2022)\citenamefont {Song}, \citenamefont {Zhang},\ and\ \citenamefont {Vishwanath}}]{song2022doping}%
  \BibitemOpen
  \bibfield  {author} {\bibinfo {author} {\bibfnamefont {X.-Y.}\ \bibnamefont {Song}}, \bibinfo {author} {\bibfnamefont {Y.-H.}\ \bibnamefont {Zhang}},\ and\ \bibinfo {author} {\bibfnamefont {A.}~\bibnamefont {Vishwanath}},\ }\bibfield  {title} {\bibinfo {title} {Doping a moir\'e mott insulator: A $t\text{\ensuremath{-}}{J}$ model study of twisted cuprates},\ }\href {https://doi.org/10.1103/PhysRevB.105.L201102} {\bibfield  {journal} {\bibinfo  {journal} {Phys. Rev. B}\ }\textbf {\bibinfo {volume} {105}},\ \bibinfo {pages} {L201102} (\bibinfo {year} {2022})}\BibitemShut {NoStop}%
\bibitem [{\citenamefont {Volkov}\ \emph {et~al.}(2023)\citenamefont {Volkov}, \citenamefont {Wilson}, \citenamefont {Lucht},\ and\ \citenamefont {Pixley}}]{volkov2023current}%
  \BibitemOpen
  \bibfield  {author} {\bibinfo {author} {\bibfnamefont {P.~A.}\ \bibnamefont {Volkov}}, \bibinfo {author} {\bibfnamefont {J.~H.}\ \bibnamefont {Wilson}}, \bibinfo {author} {\bibfnamefont {K.~P.}\ \bibnamefont {Lucht}},\ and\ \bibinfo {author} {\bibfnamefont {J.~H.}\ \bibnamefont {Pixley}},\ }\bibfield  {title} {\bibinfo {title} {Current- and field-induced topology in twisted nodal superconductors},\ }\href {https://doi.org/10.1103/PhysRevLett.130.186001} {\bibfield  {journal} {\bibinfo  {journal} {Phys. Rev. Lett.}\ }\textbf {\bibinfo {volume} {130}},\ \bibinfo {pages} {186001} (\bibinfo {year} {2023})}\BibitemShut {NoStop}%
\bibitem [{\citenamefont {Eugenio}\ and\ \citenamefont {Vafek}(2023)}]{eugenio2023twisted}%
  \BibitemOpen
  \bibfield  {author} {\bibinfo {author} {\bibfnamefont {P.~M.}\ \bibnamefont {Eugenio}}\ and\ \bibinfo {author} {\bibfnamefont {O.}~\bibnamefont {Vafek}},\ }\bibfield  {title} {\bibinfo {title} {{Twisted-bilayer FeSe and the Fe-based superlattices}},\ }\href {https://doi.org/10.21468/SciPostPhys.15.3.081} {\bibfield  {journal} {\bibinfo  {journal} {SciPost Phys.}\ }\textbf {\bibinfo {volume} {15}},\ \bibinfo {pages} {081} (\bibinfo {year} {2023})}\BibitemShut {NoStop}%
\bibitem [{\citenamefont {Kariyado}\ \emph {et~al.}(2026)\citenamefont {Kariyado}, \citenamefont {Wicaksono}, \citenamefont {Vishwanath}, \citenamefont {Volkov},\ and\ \citenamefont {Luo}}]{kariyado2026moir}%
  \BibitemOpen
  \bibfield  {author} {\bibinfo {author} {\bibfnamefont {T.}~\bibnamefont {Kariyado}}, \bibinfo {author} {\bibfnamefont {Y.}~\bibnamefont {Wicaksono}}, \bibinfo {author} {\bibfnamefont {A.}~\bibnamefont {Vishwanath}}, \bibinfo {author} {\bibfnamefont {P.}~\bibnamefont {Volkov}},\ and\ \bibinfo {author} {\bibfnamefont {Z.-X.}\ \bibnamefont {Luo}},\ }\href {https://arxiv.org/abs/2603.11153} {\bibinfo {title} {Moir\'e in $\gamma$-valley square lattice: Copper- and iron-based superconductor simulation in a single device}} (\bibinfo {year} {2026}),\ \Eprint {https://arxiv.org/abs/2603.11153} {arXiv:2603.11153 [cond-mat.str-el]} \BibitemShut {NoStop}%
\bibitem [{twi()}]{twistgamma2026}%
  \BibitemOpen
  \href@noop {} {}\bibinfo {note} {R. Shi \emph{et al}, to appear soon}\BibitemShut {NoStop}%
\bibitem [{\citenamefont {Scalapino}(2012)}]{scalapino2012}%
  \BibitemOpen
  \bibfield  {author} {\bibinfo {author} {\bibfnamefont {D.~J.}\ \bibnamefont {Scalapino}},\ }\bibfield  {title} {\bibinfo {title} {A common thread: The pairing interaction for unconventional superconductors},\ }\href {https://doi.org/10.1103/RevModPhys.84.1383} {\bibfield  {journal} {\bibinfo  {journal} {Rev. Mod. Phys.}\ }\textbf {\bibinfo {volume} {84}},\ \bibinfo {pages} {1383} (\bibinfo {year} {2012})}\BibitemShut {NoStop}%
\bibitem [{\citenamefont {Arovas}\ \emph {et~al.}(2022)\citenamefont {Arovas}, \citenamefont {Berg}, \citenamefont {Kivelson},\ and\ \citenamefont {Raghu}}]{arovas2022hubbard}%
  \BibitemOpen
  \bibfield  {author} {\bibinfo {author} {\bibfnamefont {D.~P.}\ \bibnamefont {Arovas}}, \bibinfo {author} {\bibfnamefont {E.}~\bibnamefont {Berg}}, \bibinfo {author} {\bibfnamefont {S.~A.}\ \bibnamefont {Kivelson}},\ and\ \bibinfo {author} {\bibfnamefont {S.}~\bibnamefont {Raghu}},\ }\bibfield  {title} {\bibinfo {title} {The hubbard model},\ }\href {https://doi.org/https://doi.org/10.1146/annurev-conmatphys-031620-102024} {\bibfield  {journal} {\bibinfo  {journal} {Annu. Rev. Condens. Matter Phys.}\ }\textbf {\bibinfo {volume} {13}},\ \bibinfo {pages} {239} (\bibinfo {year} {2022})}\BibitemShut {NoStop}%
\bibitem [{\citenamefont {Qin}\ \emph {et~al.}(2022)\citenamefont {Qin}, \citenamefont {Sch{\"a}fer}, \citenamefont {Andergassen}, \citenamefont {Corboz},\ and\ \citenamefont {Gull}}]{qin2022hubbard}%
  \BibitemOpen
  \bibfield  {author} {\bibinfo {author} {\bibfnamefont {M.}~\bibnamefont {Qin}}, \bibinfo {author} {\bibfnamefont {T.}~\bibnamefont {Sch{\"a}fer}}, \bibinfo {author} {\bibfnamefont {S.}~\bibnamefont {Andergassen}}, \bibinfo {author} {\bibfnamefont {P.}~\bibnamefont {Corboz}},\ and\ \bibinfo {author} {\bibfnamefont {E.}~\bibnamefont {Gull}},\ }\bibfield  {title} {\bibinfo {title} {The hubbard model: A computational perspective},\ }\href {https://doi.org/https://doi.org/10.1146/annurev-conmatphys-090921-033948} {\bibfield  {journal} {\bibinfo  {journal} {Annu. Rev. Condens. Matter Phys.}\ }\textbf {\bibinfo {volume} {13}},\ \bibinfo {pages} {275} (\bibinfo {year} {2022})}\BibitemShut {NoStop}%
\bibitem [{\citenamefont {Qu}\ \emph {et~al.}(2024)\citenamefont {Qu}, \citenamefont {Qu}, \citenamefont {Chen}, \citenamefont {Wu}, \citenamefont {Yang}, \citenamefont {Li},\ and\ \citenamefont {Su}}]{Qu2024bilayer}%
  \BibitemOpen
  \bibfield  {author} {\bibinfo {author} {\bibfnamefont {X.-Z.}\ \bibnamefont {Qu}}, \bibinfo {author} {\bibfnamefont {D.-W.}\ \bibnamefont {Qu}}, \bibinfo {author} {\bibfnamefont {J.}~\bibnamefont {Chen}}, \bibinfo {author} {\bibfnamefont {C.}~\bibnamefont {Wu}}, \bibinfo {author} {\bibfnamefont {F.}~\bibnamefont {Yang}}, \bibinfo {author} {\bibfnamefont {W.}~\bibnamefont {Li}},\ and\ \bibinfo {author} {\bibfnamefont {G.}~\bibnamefont {Su}},\ }\bibfield  {title} {\bibinfo {title} {Bilayer ${t\text{\ensuremath{-}}J\text{\ensuremath{-}}J}_{\ensuremath{\perp}}$ model and magnetically mediated pairing in the pressurized nickelate {L}a$_{3}${Ni}$_{2}${O}$_{7}$},\ }\href {https://doi.org/10.1103/PhysRevLett.132.036502} {\bibfield  {journal} {\bibinfo  {journal} {Phys. Rev. Lett.}\ }\textbf {\bibinfo {volume} {132}},\ \bibinfo {pages} {036502} (\bibinfo {year} {2024})}\BibitemShut {NoStop}%
\bibitem [{\citenamefont {Fan}\ \emph {et~al.}(2024)\citenamefont {Fan}, \citenamefont {Zhang}, \citenamefont {Zhan}, \citenamefont {Lv}, \citenamefont {Jiang}, \citenamefont {Normand},\ and\ \citenamefont {Xiang}}]{Fan2024super}%
  \BibitemOpen
  \bibfield  {author} {\bibinfo {author} {\bibfnamefont {Z.}~\bibnamefont {Fan}}, \bibinfo {author} {\bibfnamefont {J.-F.}\ \bibnamefont {Zhang}}, \bibinfo {author} {\bibfnamefont {B.}~\bibnamefont {Zhan}}, \bibinfo {author} {\bibfnamefont {D.}~\bibnamefont {Lv}}, \bibinfo {author} {\bibfnamefont {X.-Y.}\ \bibnamefont {Jiang}}, \bibinfo {author} {\bibfnamefont {B.}~\bibnamefont {Normand}},\ and\ \bibinfo {author} {\bibfnamefont {T.}~\bibnamefont {Xiang}},\ }\bibfield  {title} {\bibinfo {title} {Superconductivity in nickelate and cuprate superconductors with strong bilayer coupling},\ }\href {https://doi.org/10.1103/PhysRevB.110.024514} {\bibfield  {journal} {\bibinfo  {journal} {Phys. Rev. B}\ }\textbf {\bibinfo {volume} {110}},\ \bibinfo {pages} {024514} (\bibinfo {year} {2024})}\BibitemShut {NoStop}%
\bibitem [{\citenamefont {Kaneko}\ \emph {et~al.}(2024)\citenamefont {Kaneko}, \citenamefont {Sakakibara}, \citenamefont {Ochi},\ and\ \citenamefont {Kuroki}}]{kaneko2024pair}%
  \BibitemOpen
  \bibfield  {author} {\bibinfo {author} {\bibfnamefont {T.}~\bibnamefont {Kaneko}}, \bibinfo {author} {\bibfnamefont {H.}~\bibnamefont {Sakakibara}}, \bibinfo {author} {\bibfnamefont {M.}~\bibnamefont {Ochi}},\ and\ \bibinfo {author} {\bibfnamefont {K.}~\bibnamefont {Kuroki}},\ }\bibfield  {title} {\bibinfo {title} {Pair correlations in the two-orbital hubbard ladder: Implications for superconductivity in the bilayer nickelate {L}a$_{3}${N}i$_{2}${O}$_{7}$},\ }\href {https://doi.org/10.1103/PhysRevB.109.045154} {\bibfield  {journal} {\bibinfo  {journal} {Phys. Rev. B}\ }\textbf {\bibinfo {volume} {109}},\ \bibinfo {pages} {045154} (\bibinfo {year} {2024})}\BibitemShut {NoStop}%
\bibitem [{\citenamefont {Lu}\ \emph {et~al.}(2024{\natexlab{b}})\citenamefont {Lu}, \citenamefont {Pan}, \citenamefont {Yang},\ and\ \citenamefont {Wu}}]{lu2024interlayer}%
  \BibitemOpen
  \bibfield  {author} {\bibinfo {author} {\bibfnamefont {C.}~\bibnamefont {Lu}}, \bibinfo {author} {\bibfnamefont {Z.}~\bibnamefont {Pan}}, \bibinfo {author} {\bibfnamefont {F.}~\bibnamefont {Yang}},\ and\ \bibinfo {author} {\bibfnamefont {C.}~\bibnamefont {Wu}},\ }\bibfield  {title} {\bibinfo {title} {Interlayer-coupling-driven high-temperature superconductivity in {L}a$_{3}${N}i$_{2}${O}$_{7}$ under pressure},\ }\href {https://doi.org/10.1103/PhysRevLett.132.146002} {\bibfield  {journal} {\bibinfo  {journal} {Phys. Rev. Lett.}\ }\textbf {\bibinfo {volume} {132}},\ \bibinfo {pages} {146002} (\bibinfo {year} {2024}{\natexlab{b}})}\BibitemShut {NoStop}%
\bibitem [{\citenamefont {Jiang}\ \emph {et~al.}(2025)\citenamefont {Jiang}, \citenamefont {Cao}, \citenamefont {Yang}, \citenamefont {Lu},\ and\ \citenamefont {Wang}}]{jiang2025theory}%
  \BibitemOpen
  \bibfield  {author} {\bibinfo {author} {\bibfnamefont {K.-Y.}\ \bibnamefont {Jiang}}, \bibinfo {author} {\bibfnamefont {Y.-H.}\ \bibnamefont {Cao}}, \bibinfo {author} {\bibfnamefont {Q.-G.}\ \bibnamefont {Yang}}, \bibinfo {author} {\bibfnamefont {H.-Y.}\ \bibnamefont {Lu}},\ and\ \bibinfo {author} {\bibfnamefont {Q.-H.}\ \bibnamefont {Wang}},\ }\bibfield  {title} {\bibinfo {title} {Theory of pressure dependence of superconductivity in bilayer nickelate {L}a$_{3}${N}i$_{2}${O}$_{7}$},\ }\href {https://doi.org/10.1103/PhysRevLett.134.076001} {\bibfield  {journal} {\bibinfo  {journal} {Phys. Rev. Lett.}\ }\textbf {\bibinfo {volume} {134}},\ \bibinfo {pages} {076001} (\bibinfo {year} {2025})}\BibitemShut {NoStop}%
\bibitem [{\citenamefont {Zhang}\ \emph {et~al.}(2024)\citenamefont {Zhang}, \citenamefont {Wang}, \citenamefont {Liu}, \citenamefont {Fan}, \citenamefont {Cao},\ and\ \citenamefont {Xiao}}]{zhang2024polarization}%
  \BibitemOpen
  \bibfield  {author} {\bibinfo {author} {\bibfnamefont {X.-W.}\ \bibnamefont {Zhang}}, \bibinfo {author} {\bibfnamefont {C.}~\bibnamefont {Wang}}, \bibinfo {author} {\bibfnamefont {X.}~\bibnamefont {Liu}}, \bibinfo {author} {\bibfnamefont {Y.}~\bibnamefont {Fan}}, \bibinfo {author} {\bibfnamefont {T.}~\bibnamefont {Cao}},\ and\ \bibinfo {author} {\bibfnamefont {D.}~\bibnamefont {Xiao}},\ }\bibfield  {title} {\bibinfo {title} {Polarization-driven band topology evolution in twisted {M}o{T}e$_2$ and {WS}e$_2$},\ }\href {https://doi.org/https://doi.org/10.1038/s41467-024-48511-x} {\bibfield  {journal} {\bibinfo  {journal} {Nat. Commun.}\ }\textbf {\bibinfo {volume} {15}},\ \bibinfo {pages} {4223} (\bibinfo {year} {2024})}\BibitemShut {NoStop}%
\bibitem [{\citenamefont {Li}\ and\ \citenamefont {Wu}(2017)}]{li2017binary}%
  \BibitemOpen
  \bibfield  {author} {\bibinfo {author} {\bibfnamefont {L.}~\bibnamefont {Li}}\ and\ \bibinfo {author} {\bibfnamefont {M.}~\bibnamefont {Wu}},\ }\bibfield  {title} {\bibinfo {title} {Binary compound bilayer and multilayer with vertical polarizations: two-dimensional ferroelectrics, multiferroics, and nanogenerators},\ }\href {https://doi.org/https://doi.org/10.1021/acsnano.7b02756} {\bibfield  {journal} {\bibinfo  {journal} {ACS Nano}\ }\textbf {\bibinfo {volume} {11}},\ \bibinfo {pages} {6382} (\bibinfo {year} {2017})}\BibitemShut {NoStop}%
\bibitem [{\citenamefont {Yasuda}\ \emph {et~al.}(2021)\citenamefont {Yasuda}, \citenamefont {Wang}, \citenamefont {Watanabe}, \citenamefont {Taniguchi},\ and\ \citenamefont {Jarillo-Herrero}}]{yasuda2021stacking}%
  \BibitemOpen
  \bibfield  {author} {\bibinfo {author} {\bibfnamefont {K.}~\bibnamefont {Yasuda}}, \bibinfo {author} {\bibfnamefont {X.}~\bibnamefont {Wang}}, \bibinfo {author} {\bibfnamefont {K.}~\bibnamefont {Watanabe}}, \bibinfo {author} {\bibfnamefont {T.}~\bibnamefont {Taniguchi}},\ and\ \bibinfo {author} {\bibfnamefont {P.}~\bibnamefont {Jarillo-Herrero}},\ }\bibfield  {title} {\bibinfo {title} {Stacking-engineered ferroelectricity in bilayer boron nitride},\ }\href {https://doi.org/https://doi.org/10.1126/science.abd3230} {\bibfield  {journal} {\bibinfo  {journal} {Science}\ }\textbf {\bibinfo {volume} {372}},\ \bibinfo {pages} {1458} (\bibinfo {year} {2021})}\BibitemShut {NoStop}%
\bibitem [{\citenamefont {Vizner~Stern}\ \emph {et~al.}(2021)\citenamefont {Vizner~Stern}, \citenamefont {Waschitz}, \citenamefont {Cao}, \citenamefont {Nevo}, \citenamefont {Watanabe}, \citenamefont {Taniguchi}, \citenamefont {Sela}, \citenamefont {Urbakh}, \citenamefont {Hod},\ and\ \citenamefont {Ben~Shalom}}]{vizner2021interfacial}%
  \BibitemOpen
  \bibfield  {author} {\bibinfo {author} {\bibfnamefont {M.}~\bibnamefont {Vizner~Stern}}, \bibinfo {author} {\bibfnamefont {Y.}~\bibnamefont {Waschitz}}, \bibinfo {author} {\bibfnamefont {W.}~\bibnamefont {Cao}}, \bibinfo {author} {\bibfnamefont {I.}~\bibnamefont {Nevo}}, \bibinfo {author} {\bibfnamefont {K.}~\bibnamefont {Watanabe}}, \bibinfo {author} {\bibfnamefont {T.}~\bibnamefont {Taniguchi}}, \bibinfo {author} {\bibfnamefont {E.}~\bibnamefont {Sela}}, \bibinfo {author} {\bibfnamefont {M.}~\bibnamefont {Urbakh}}, \bibinfo {author} {\bibfnamefont {O.}~\bibnamefont {Hod}},\ and\ \bibinfo {author} {\bibfnamefont {M.}~\bibnamefont {Ben~Shalom}},\ }\bibfield  {title} {\bibinfo {title} {Interfacial ferroelectricity by van der waals sliding},\ }\href {https://doi.org/https://doi.org/10.1126/science.abe8177} {\bibfield  {journal} {\bibinfo  {journal} {Science}\ }\textbf {\bibinfo {volume} {372}},\ \bibinfo {pages} {1462} (\bibinfo {year} {2021})}\BibitemShut {NoStop}%
\bibitem [{\citenamefont {Fei}\ \emph {et~al.}(2018)\citenamefont {Fei}, \citenamefont {Zhao}, \citenamefont {Palomaki}, \citenamefont {Sun}, \citenamefont {Miller}, \citenamefont {Zhao}, \citenamefont {Yan}, \citenamefont {Xu},\ and\ \citenamefont {Cobden}}]{fei2018ferroelectric}%
  \BibitemOpen
  \bibfield  {author} {\bibinfo {author} {\bibfnamefont {Z.}~\bibnamefont {Fei}}, \bibinfo {author} {\bibfnamefont {W.}~\bibnamefont {Zhao}}, \bibinfo {author} {\bibfnamefont {T.~A.}\ \bibnamefont {Palomaki}}, \bibinfo {author} {\bibfnamefont {B.}~\bibnamefont {Sun}}, \bibinfo {author} {\bibfnamefont {M.~K.}\ \bibnamefont {Miller}}, \bibinfo {author} {\bibfnamefont {Z.}~\bibnamefont {Zhao}}, \bibinfo {author} {\bibfnamefont {J.}~\bibnamefont {Yan}}, \bibinfo {author} {\bibfnamefont {X.}~\bibnamefont {Xu}},\ and\ \bibinfo {author} {\bibfnamefont {D.~H.}\ \bibnamefont {Cobden}},\ }\bibfield  {title} {\bibinfo {title} {Ferroelectric switching of a two-dimensional metal},\ }\href {https://doi.org/https://doi.org/10.1038/s41586-018-0336-3} {\bibfield  {journal} {\bibinfo  {journal} {Nature}\ }\textbf {\bibinfo {volume} {560}},\ \bibinfo {pages} {336} (\bibinfo {year} {2018})}\BibitemShut {NoStop}%
\bibitem [{\citenamefont {Wang}\ \emph {et~al.}(2022)\citenamefont {Wang}, \citenamefont {Yasuda}, \citenamefont {Zhang}, \citenamefont {Liu}, \citenamefont {Watanabe}, \citenamefont {Taniguchi}, \citenamefont {Hone}, \citenamefont {Fu},\ and\ \citenamefont {Jarillo-Herrero}}]{wang2022interfacial}%
  \BibitemOpen
  \bibfield  {author} {\bibinfo {author} {\bibfnamefont {X.}~\bibnamefont {Wang}}, \bibinfo {author} {\bibfnamefont {K.}~\bibnamefont {Yasuda}}, \bibinfo {author} {\bibfnamefont {Y.}~\bibnamefont {Zhang}}, \bibinfo {author} {\bibfnamefont {S.}~\bibnamefont {Liu}}, \bibinfo {author} {\bibfnamefont {K.}~\bibnamefont {Watanabe}}, \bibinfo {author} {\bibfnamefont {T.}~\bibnamefont {Taniguchi}}, \bibinfo {author} {\bibfnamefont {J.}~\bibnamefont {Hone}}, \bibinfo {author} {\bibfnamefont {L.}~\bibnamefont {Fu}},\ and\ \bibinfo {author} {\bibfnamefont {P.}~\bibnamefont {Jarillo-Herrero}},\ }\bibfield  {title} {\bibinfo {title} {Interfacial ferroelectricity in rhombohedral-stacked bilayer transition metal dichalcogenides},\ }\href {https://doi.org/https://doi.org/10.1038/s41565-021-01059-z} {\bibfield  {journal} {\bibinfo  {journal} {Nat. Nanotechnol.}\ }\textbf {\bibinfo {volume} {17}},\ \bibinfo {pages} {367} (\bibinfo {year} {2022})}\BibitemShut {NoStop}%
\bibitem [{\citenamefont {Wan}\ \emph {et~al.}(2022)\citenamefont {Wan}, \citenamefont {Hu}, \citenamefont {Mao}, \citenamefont {Fu}, \citenamefont {Yuan}, \citenamefont {Song}, \citenamefont {Gan}, \citenamefont {Xu}, \citenamefont {Xue}, \citenamefont {Cheng}, \citenamefont {Huang}, \citenamefont {Yang}, \citenamefont {Dai}, \citenamefont {Zeng},\ and\ \citenamefont {Kan}}]{Wan2022room}%
  \BibitemOpen
  \bibfield  {author} {\bibinfo {author} {\bibfnamefont {Y.}~\bibnamefont {Wan}}, \bibinfo {author} {\bibfnamefont {T.}~\bibnamefont {Hu}}, \bibinfo {author} {\bibfnamefont {X.}~\bibnamefont {Mao}}, \bibinfo {author} {\bibfnamefont {J.}~\bibnamefont {Fu}}, \bibinfo {author} {\bibfnamefont {K.}~\bibnamefont {Yuan}}, \bibinfo {author} {\bibfnamefont {Y.}~\bibnamefont {Song}}, \bibinfo {author} {\bibfnamefont {X.}~\bibnamefont {Gan}}, \bibinfo {author} {\bibfnamefont {X.}~\bibnamefont {Xu}}, \bibinfo {author} {\bibfnamefont {M.}~\bibnamefont {Xue}}, \bibinfo {author} {\bibfnamefont {X.}~\bibnamefont {Cheng}}, \bibinfo {author} {\bibfnamefont {C.}~\bibnamefont {Huang}}, \bibinfo {author} {\bibfnamefont {J.}~\bibnamefont {Yang}}, \bibinfo {author} {\bibfnamefont {L.}~\bibnamefont {Dai}}, \bibinfo {author} {\bibfnamefont {H.}~\bibnamefont {Zeng}},\ and\ \bibinfo {author} {\bibfnamefont {E.}~\bibnamefont {Kan}},\ }\bibfield  {title} {\bibinfo {title} {Room-temperature ferroelectricity in 1{T}$'$-{R}e{S}$_{2}$
  multilayers},\ }\href {https://doi.org/10.1103/PhysRevLett.128.067601} {\bibfield  {journal} {\bibinfo  {journal} {Phys. Rev. Lett.}\ }\textbf {\bibinfo {volume} {128}},\ \bibinfo {pages} {067601} (\bibinfo {year} {2022})}\BibitemShut {NoStop}%
\bibitem [{\citenamefont {Ji}\ \emph {et~al.}(2023)\citenamefont {Ji}, \citenamefont {Yu}, \citenamefont {Xu},\ and\ \citenamefont {Xiang}}]{ji2023general}%
  \BibitemOpen
  \bibfield  {author} {\bibinfo {author} {\bibfnamefont {J.}~\bibnamefont {Ji}}, \bibinfo {author} {\bibfnamefont {G.}~\bibnamefont {Yu}}, \bibinfo {author} {\bibfnamefont {C.}~\bibnamefont {Xu}},\ and\ \bibinfo {author} {\bibfnamefont {H.~J.}\ \bibnamefont {Xiang}},\ }\bibfield  {title} {\bibinfo {title} {General theory for bilayer stacking ferroelectricity},\ }\href {https://doi.org/10.1103/PhysRevLett.130.146801} {\bibfield  {journal} {\bibinfo  {journal} {Phys. Rev. Lett.}\ }\textbf {\bibinfo {volume} {130}},\ \bibinfo {pages} {146801} (\bibinfo {year} {2023})}\BibitemShut {NoStop}%
\bibitem [{\citenamefont {Yang}\ \emph {et~al.}(2023)\citenamefont {Yang}, \citenamefont {Ding}, \citenamefont {Gao},\ and\ \citenamefont {Wu}}]{liu2023atypical}%
  \BibitemOpen
  \bibfield  {author} {\bibinfo {author} {\bibfnamefont {L.}~\bibnamefont {Yang}}, \bibinfo {author} {\bibfnamefont {S.}~\bibnamefont {Ding}}, \bibinfo {author} {\bibfnamefont {J.}~\bibnamefont {Gao}},\ and\ \bibinfo {author} {\bibfnamefont {M.}~\bibnamefont {Wu}},\ }\bibfield  {title} {\bibinfo {title} {Atypical sliding and moir\'e ferroelectricity in pure multilayer graphene},\ }\href {https://doi.org/10.1103/PhysRevLett.131.096801} {\bibfield  {journal} {\bibinfo  {journal} {Phys. Rev. Lett.}\ }\textbf {\bibinfo {volume} {131}},\ \bibinfo {pages} {096801} (\bibinfo {year} {2023})}\BibitemShut {NoStop}%
\bibitem [{sup()}]{supple}%
  \BibitemOpen
  \href@noop {} {}\bibinfo {note} {See Supplemental Material for details}\BibitemShut {NoStop}%
\bibitem [{\citenamefont {Akashi}\ \emph {et~al.}(2017)\citenamefont {Akashi}, \citenamefont {Iida}, \citenamefont {Yamamoto},\ and\ \citenamefont {Yoshizawa}}]{Akashi2017}%
  \BibitemOpen
  \bibfield  {author} {\bibinfo {author} {\bibfnamefont {R.}~\bibnamefont {Akashi}}, \bibinfo {author} {\bibfnamefont {Y.}~\bibnamefont {Iida}}, \bibinfo {author} {\bibfnamefont {K.}~\bibnamefont {Yamamoto}},\ and\ \bibinfo {author} {\bibfnamefont {K.}~\bibnamefont {Yoshizawa}},\ }\bibfield  {title} {\bibinfo {title} {Interference of the bloch phase in layered materials with stacking shifts},\ }\href {https://doi.org/10.1103/PhysRevB.95.245401} {\bibfield  {journal} {\bibinfo  {journal} {Phys. Rev. B}\ }\textbf {\bibinfo {volume} {95}},\ \bibinfo {pages} {245401} (\bibinfo {year} {2017})}\BibitemShut {NoStop}%
\bibitem [{\citenamefont {Chen}\ \emph {et~al.}(2022)\citenamefont {Chen}, \citenamefont {Yang},\ and\ \citenamefont {Zhao}}]{Chen2022}%
  \BibitemOpen
  \bibfield  {author} {\bibinfo {author} {\bibfnamefont {Z.~Y.}\ \bibnamefont {Chen}}, \bibinfo {author} {\bibfnamefont {S.~A.}\ \bibnamefont {Yang}},\ and\ \bibinfo {author} {\bibfnamefont {Y.~X.}\ \bibnamefont {Zhao}},\ }\bibfield  {title} {\bibinfo {title} {Brillouin klein bottle from artificial gauge fields},\ }\href {https://doi.org/10.1038/s41467-022-29953-7} {\bibfield  {journal} {\bibinfo  {journal} {Nat. Commun.}\ }\textbf {\bibinfo {volume} {13}},\ \bibinfo {pages} {2215} (\bibinfo {year} {2022})}\BibitemShut {NoStop}%
\bibitem [{\citenamefont {Zhang}\ \emph {et~al.}(2023)\citenamefont {Zhang}, \citenamefont {Chen}, \citenamefont {Zhang},\ and\ \citenamefont {Zhao}}]{Zhang2023}%
  \BibitemOpen
  \bibfield  {author} {\bibinfo {author} {\bibfnamefont {C.}~\bibnamefont {Zhang}}, \bibinfo {author} {\bibfnamefont {Z.~Y.}\ \bibnamefont {Chen}}, \bibinfo {author} {\bibfnamefont {Z.}~\bibnamefont {Zhang}},\ and\ \bibinfo {author} {\bibfnamefont {Y.~X.}\ \bibnamefont {Zhao}},\ }\bibfield  {title} {\bibinfo {title} {General theory of momentum-space nonsymmorphic symmetry},\ }\href {https://doi.org/10.1103/PhysRevLett.130.256601} {\bibfield  {journal} {\bibinfo  {journal} {Phys. Rev. Lett.}\ }\textbf {\bibinfo {volume} {130}},\ \bibinfo {pages} {256601} (\bibinfo {year} {2023})}\BibitemShut {NoStop}%
\bibitem [{\citenamefont {Pruss}\ \emph {et~al.}(1993)\citenamefont {Pruss}, \citenamefont {Snyder},\ and\ \citenamefont {Stacy}}]{pruss1993new}%
  \BibitemOpen
  \bibfield  {author} {\bibinfo {author} {\bibfnamefont {E.~A.}\ \bibnamefont {Pruss}}, \bibinfo {author} {\bibfnamefont {B.~S.}\ \bibnamefont {Snyder}},\ and\ \bibinfo {author} {\bibfnamefont {A.~M.}\ \bibnamefont {Stacy}},\ }\bibfield  {title} {\bibinfo {title} {A new layered ternary sulfide: Formation of {C}u$_2${WS}$_4$ by reaction of ws and {C}u$^+$ ions},\ }\href {https://doi.org/10.1002/anie.199302561} {\bibfield  {journal} {\bibinfo  {journal} {Angew. Chem., Int. Ed. Engl.}\ }\textbf {\bibinfo {volume} {32}},\ \bibinfo {pages} {256} (\bibinfo {year} {1993})}\BibitemShut {NoStop}%
\bibitem [{\citenamefont {Crossland}\ \emph {et~al.}(2005)\citenamefont {Crossland}, \citenamefont {Hickey},\ and\ \citenamefont {Evans}}]{crossland2005synthesis}%
  \BibitemOpen
  \bibfield  {author} {\bibinfo {author} {\bibfnamefont {C.~J.}\ \bibnamefont {Crossland}}, \bibinfo {author} {\bibfnamefont {P.~J.}\ \bibnamefont {Hickey}},\ and\ \bibinfo {author} {\bibfnamefont {J.~S.}\ \bibnamefont {Evans}},\ }\bibfield  {title} {\bibinfo {title} {The synthesis and characterisation of {C}u$_2${MX}e$_4$ ({M}= {W} or {M}o; {X}= {S}, {S}e or {S}/{S}e) materials prepared by a solvothermal method},\ }\href {https://doi.org/https://doi.org/10.1039/B507129A} {\bibfield  {journal} {\bibinfo  {journal} {J. Mater. Chem.}\ }\textbf {\bibinfo {volume} {15}},\ \bibinfo {pages} {3452} (\bibinfo {year} {2005})}\BibitemShut {NoStop}%
\bibitem [{\citenamefont {Chen}\ \emph {et~al.}(2014)\citenamefont {Chen}, \citenamefont {Chen}, \citenamefont {Zhu}, \citenamefont {Gao}, \citenamefont {Luo}, \citenamefont {Wang}, \citenamefont {Zhang}, \citenamefont {Zhang}, \citenamefont {Wang}, \citenamefont {Xiong} \emph {et~al.}}]{chen2014solvothermal}%
  \BibitemOpen
  \bibfield  {author} {\bibinfo {author} {\bibfnamefont {W.}~\bibnamefont {Chen}}, \bibinfo {author} {\bibfnamefont {H.}~\bibnamefont {Chen}}, \bibinfo {author} {\bibfnamefont {H.}~\bibnamefont {Zhu}}, \bibinfo {author} {\bibfnamefont {Q.}~\bibnamefont {Gao}}, \bibinfo {author} {\bibfnamefont {J.}~\bibnamefont {Luo}}, \bibinfo {author} {\bibfnamefont {Y.}~\bibnamefont {Wang}}, \bibinfo {author} {\bibfnamefont {S.}~\bibnamefont {Zhang}}, \bibinfo {author} {\bibfnamefont {K.}~\bibnamefont {Zhang}}, \bibinfo {author} {\bibfnamefont {C.}~\bibnamefont {Wang}}, \bibinfo {author} {\bibfnamefont {Y.}~\bibnamefont {Xiong}}, \emph {et~al.},\ }\bibfield  {title} {\bibinfo {title} {Solvothermal synthesis of ternary {C}u$_2${M}o{S}$_4$ nanosheets: structural characterization at the atomic level},\ }\href {https://doi.org/https://doi.org/10.1002/smll.201400752} {\bibfield  {journal} {\bibinfo  {journal} {Small}\ }\textbf {\bibinfo {volume} {10}},\ \bibinfo {pages} {4637} (\bibinfo {year} {2014})}\BibitemShut {NoStop}%
\bibitem [{\citenamefont {Lin}\ \emph {et~al.}(2019)\citenamefont {Lin}, \citenamefont {Chen}, \citenamefont {Zhang},\ and\ \citenamefont {Song}}]{lin2019recent}%
  \BibitemOpen
  \bibfield  {author} {\bibinfo {author} {\bibfnamefont {Y.}~\bibnamefont {Lin}}, \bibinfo {author} {\bibfnamefont {S.}~\bibnamefont {Chen}}, \bibinfo {author} {\bibfnamefont {K.}~\bibnamefont {Zhang}},\ and\ \bibinfo {author} {\bibfnamefont {L.}~\bibnamefont {Song}},\ }\bibfield  {title} {\bibinfo {title} {Recent advances of ternary layered {C}u$_2${MX}$_4$ ({M}= {W} or {M}o; {X}= {S}, {S}e or {S}/{S}e) nanomaterials for photocatalysis},\ }\href {https://doi.org/https://doi.org/10.1002/solr.201800320} {\bibfield  {journal} {\bibinfo  {journal} {Solar RRL}\ }\textbf {\bibinfo {volume} {3}},\ \bibinfo {pages} {1800320} (\bibinfo {year} {2019})}\BibitemShut {NoStop}%
\bibitem [{\citenamefont {Jiang}\ \emph {et~al.}(2024)\citenamefont {Jiang}, \citenamefont {Petralanda}, \citenamefont {Skorupskii}, \citenamefont {Xu}, \citenamefont {Pi}, \citenamefont {C{\u{a}}lug{\u{a}}ru}, \citenamefont {Hu}, \citenamefont {Xie}, \citenamefont {Mustaf}, \citenamefont {H{\"o}hn} \emph {et~al.}}]{jiang20242d}%
  \BibitemOpen
  \bibfield  {author} {\bibinfo {author} {\bibfnamefont {Y.}~\bibnamefont {Jiang}}, \bibinfo {author} {\bibfnamefont {U.}~\bibnamefont {Petralanda}}, \bibinfo {author} {\bibfnamefont {G.}~\bibnamefont {Skorupskii}}, \bibinfo {author} {\bibfnamefont {Q.}~\bibnamefont {Xu}}, \bibinfo {author} {\bibfnamefont {H.}~\bibnamefont {Pi}}, \bibinfo {author} {\bibfnamefont {D.}~\bibnamefont {C{\u{a}}lug{\u{a}}ru}}, \bibinfo {author} {\bibfnamefont {H.}~\bibnamefont {Hu}}, \bibinfo {author} {\bibfnamefont {J.}~\bibnamefont {Xie}}, \bibinfo {author} {\bibfnamefont {R.~A.}\ \bibnamefont {Mustaf}}, \bibinfo {author} {\bibfnamefont {P.}~\bibnamefont {H{\"o}hn}}, \emph {et~al.},\ }\href {https://arxiv.org/abs/2411.09741} {\bibinfo {title} {2d theoretically twistable material database}} (\bibinfo {year} {2024}),\ \Eprint {https://arxiv.org/abs/2411.09741} {arXiv:2411.09741 [cond-mat.mtrl-sci]} \BibitemShut {NoStop}%
\bibitem [{\citenamefont {Bouadim}\ \emph {et~al.}(2008)\citenamefont {Bouadim}, \citenamefont {Batrouni}, \citenamefont {H{\'e}bert},\ and\ \citenamefont {Scalettar}}]{bouadim2008}%
  \BibitemOpen
  \bibfield  {author} {\bibinfo {author} {\bibfnamefont {K.}~\bibnamefont {Bouadim}}, \bibinfo {author} {\bibfnamefont {G.~G.}\ \bibnamefont {Batrouni}}, \bibinfo {author} {\bibfnamefont {F.}~\bibnamefont {H{\'e}bert}},\ and\ \bibinfo {author} {\bibfnamefont {R.~T.}\ \bibnamefont {Scalettar}},\ }\bibfield  {title} {\bibinfo {title} {Magnetic and transport properties of a coupled hubbard bilayer with electron and hole doping},\ }\href {https://doi.org/10.1103/PhysRevB.77.144527} {\bibfield  {journal} {\bibinfo  {journal} {Phys. Rev. B}\ }\textbf {\bibinfo {volume} {77}},\ \bibinfo {pages} {144527} (\bibinfo {year} {2008})}\BibitemShut {NoStop}%
\bibitem [{\citenamefont {Mou}\ \emph {et~al.}(2022)\citenamefont {Mou}, \citenamefont {Mondaini},\ and\ \citenamefont {Scalettar}}]{mou2022}%
  \BibitemOpen
  \bibfield  {author} {\bibinfo {author} {\bibfnamefont {Y.}~\bibnamefont {Mou}}, \bibinfo {author} {\bibfnamefont {R.}~\bibnamefont {Mondaini}},\ and\ \bibinfo {author} {\bibfnamefont {R.~T.}\ \bibnamefont {Scalettar}},\ }\bibfield  {title} {\bibinfo {title} {Bilayer hubbard model: Analysis based on the fermionic sign problem},\ }\href {https://doi.org/10.1103/PhysRevB.106.125116} {\bibfield  {journal} {\bibinfo  {journal} {Phys. Rev. B}\ }\textbf {\bibinfo {volume} {106}},\ \bibinfo {pages} {125116} (\bibinfo {year} {2022})}\BibitemShut {NoStop}%
\bibitem [{\citenamefont {Bulut}\ \emph {et~al.}(1992)\citenamefont {Bulut}, \citenamefont {Scalapino},\ and\ \citenamefont {Scalettar}}]{bulut1992}%
  \BibitemOpen
  \bibfield  {author} {\bibinfo {author} {\bibfnamefont {N.}~\bibnamefont {Bulut}}, \bibinfo {author} {\bibfnamefont {D.~J.}\ \bibnamefont {Scalapino}},\ and\ \bibinfo {author} {\bibfnamefont {R.~T.}\ \bibnamefont {Scalettar}},\ }\bibfield  {title} {\bibinfo {title} {Nodeless $d$-wave pairing in a two-layer hubbard model},\ }\href {https://doi.org/10.1103/PhysRevB.45.5577} {\bibfield  {journal} {\bibinfo  {journal} {Phys. Rev. B}\ }\textbf {\bibinfo {volume} {45}},\ \bibinfo {pages} {5577} (\bibinfo {year} {1992})}\BibitemShut {NoStop}%
\bibitem [{\citenamefont {Maier}\ and\ \citenamefont {Scalapino}(2011)}]{maier2011}%
  \BibitemOpen
  \bibfield  {author} {\bibinfo {author} {\bibfnamefont {T.~A.}\ \bibnamefont {Maier}}\ and\ \bibinfo {author} {\bibfnamefont {D.~J.}\ \bibnamefont {Scalapino}},\ }\bibfield  {title} {\bibinfo {title} {Pair structure and the pairing interaction in a bilayer hubbard model for unconventional superconductivity},\ }\href {https://doi.org/10.1103/PhysRevB.84.180513} {\bibfield  {journal} {\bibinfo  {journal} {Phys. Rev. B}\ }\textbf {\bibinfo {volume} {84}},\ \bibinfo {pages} {180513} (\bibinfo {year} {2011})}\BibitemShut {NoStop}%
\bibitem [{\citenamefont {Normand}\ and\ \citenamefont {Kampf}(2001)}]{normand2001}%
  \BibitemOpen
  \bibfield  {author} {\bibinfo {author} {\bibfnamefont {B.}~\bibnamefont {Normand}}\ and\ \bibinfo {author} {\bibfnamefont {A.~P.}\ \bibnamefont {Kampf}},\ }\bibfield  {title} {\bibinfo {title} {Lattice anisotropy as the microscopic origin of static stripes in cuprates},\ }\href {https://doi.org/10.1103/PhysRevB.64.024521} {\bibfield  {journal} {\bibinfo  {journal} {Phys. Rev. B}\ }\textbf {\bibinfo {volume} {64}},\ \bibinfo {pages} {024521} (\bibinfo {year} {2001})}\BibitemShut {NoStop}%
\bibitem [{\citenamefont {Vaidya}\ \emph {et~al.}(2025)\citenamefont {Vaidya}, \citenamefont {Fonseca}, \citenamefont {Hirsbrunner}, \citenamefont {Hughes},\ and\ \citenamefont {Solja\ifmmode \check{c}\else \v{c}\fi{}i\ifmmode~\acute{c}\else \'{c}\fi{}}}]{Vaidya2025}%
  \BibitemOpen
  \bibfield  {author} {\bibinfo {author} {\bibfnamefont {S.}~\bibnamefont {Vaidya}}, \bibinfo {author} {\bibfnamefont {A.~G.}\ \bibnamefont {Fonseca}}, \bibinfo {author} {\bibfnamefont {M.~R.}\ \bibnamefont {Hirsbrunner}}, \bibinfo {author} {\bibfnamefont {T.~L.}\ \bibnamefont {Hughes}},\ and\ \bibinfo {author} {\bibfnamefont {M.}~\bibnamefont {Solja\ifmmode \check{c}\else \v{c}\fi{}i\ifmmode~\acute{c}\else \'{c}\fi{}}},\ }\bibfield  {title} {\bibinfo {title} {Quantized crystalline-electromagnetic responses in insulators},\ }\href {https://doi.org/10.1103/5nsr-rw4d} {\bibfield  {journal} {\bibinfo  {journal} {Phys. Rev. Lett.}\ }\textbf {\bibinfo {volume} {135}},\ \bibinfo {pages} {256602} (\bibinfo {year} {2025})}\BibitemShut {NoStop}%
\end{thebibliography}

%

\end{document}